\documentclass[onecolumn,12pt,draftcls]{IEEEtran}
\IEEEoverridecommandlockouts
\usepackage{amsmath}
\usepackage{amsfonts}
\usepackage{bbding}
\usepackage{amssymb}
\usepackage{array}
\usepackage{subfigure}
\usepackage{textcomp}

\usepackage{graphicx}
\usepackage{subfigure}
\usepackage[named]{algo}
\usepackage{algorithmic}
\usepackage{psfrag}
\usepackage{stfloats}
\usepackage[compress]{cite}
\makeatletter
\renewcommand{\citepunct}{,\penalty\@m\hskip.13emplus.1emminus.1em}
\renewcommand{\citedash}{\hbox{--}\penalty\@m}
\makeatother
\usepackage{setspace}
\usepackage{color}
\allowdisplaybreaks

\usepackage{amsthm}
\usepackage{stfloats}

\newtheorem{rem}{Remark}
\newtheorem{pro}{Property}
\newtheorem{prop}{Proposition}

\usepackage{hyperref}
\usepackage{bm}
\renewcommand{\IEEEQED}{\IEEEQEDopen}

\begin{document}
\title{Energy Efficient Resource Allocation for Hybrid Services with Future Channel Gains}

\author{\IEEEauthorblockN{Changyang She and Chenyang Yang}\thanks{A part of this work was presented in IEEE/CIC ICCC 2015 \cite{ICCC2015}.}
%
}

\maketitle
\vspace{-15mm}\begin{abstract}
In this paper, we propose a framework to maximize energy efficiency (EE) of a system supporting real-time (RT) and non-real-time services by exploiting future average channel gains of mobile users, which change in the timescale of seconds and are reported predictable within a minute-long time window. To demonstrate the potential of improving EE by jointly optimizing resource allocation for both services by harnessing both future average channel gains and current instantaneous channel gains, we optimize a two-timescale policy with perfect prediction, by taking orthogonal frequency division multiple access system serving RT and video-on-demand (VoD) users as an example. Considering that fine-grained prediction for every user is with high cost, we propose a heuristic policy that only needs to predict the median of average channel gains of VoD users. Simulation results show that the optimal policy outperforms relevant counterparts, indicating the necessity of the joint optimization for both services and for two timescales. Besides, the heuristic policy performs closely to the optimal policy with perfect prediction while becomes superior with large prediction errors. This suggests that the EE gain over non-predictive policies can be captured with coarse-grained prediction.
\end{abstract}

\vspace{-6mm}\begin{IEEEkeywords}\vspace{-3mm}
Energy efficiency, predictive resource allocation, future information, VoD services, real-time services
\end{IEEEkeywords}

\vspace{-10mm}\section{Introduction}
To support the ever-growing traffic demands, the main trend techniques of future mobile communications are exploring wider spectrum and deploying more antennas or base stations (BSs). Yet due to temporal-spatial traffic fluctuations, existing cellular networks, usually optimized for fully loaded scenarios, are often observed not-busy in many places.
According to the recent data analysis in \cite{NB2018}, only a small portion of radio resources in Long Term Evolution (LTE) networks are truly used in average. To avoid wasting resources when the system is not fully loaded, energy efficiency (EE) becomes a key performance metric for cellular networks \cite{Shunqing2017Fundamental}. Recently, energy-efficient resource allocation has been investigated extensively in literature \cite{EEECLY,Amir2013Energy,Changyang2015Tcom,yang2018energy,xu2018energy}.

The dynamic nature of traffic load comes from user behaviors such as mobility and activities, which change in a much longer timescale than channel state information (CSI) and have long been regarded as random in wireless system design.
However, the research efforts in other domains demonstrate that some user behaviors, say mobility pattern, are highly predictable \cite{Apollinaire2015A,LSTM_highway,CGZ2018ICC,bui2016anticipatory}. With the predicted trajectory, radio resources can be allocated adaptive to network dynamics caused by user mobility. This provides a promising way to circumvent the resource under-utilization. By harnessing future information in a minute-level time horizon, predictive resource allocation (PRA) has been shown to provide remarkable gain in terms of improving network EE, throughput, and user experience than the non-predictive counterparts \cite{Hatem2014EE,Apollinaire2015Mobility,DA2016ICC,Robert2014INFOCOM,Yao2016Planning,bui2016anticipatory,Ramy2017TWC,Ramy2018GCT,guo2018icc,guo2018interference}. The gain of PRA has been validated by recent data-driven analysis \cite{Robert2014INFOCOM,Suhail2017ICC,NB2018}.

Prevalent resource allocation policies are non-predictive, which
are optimized with instantaneous or average channel gain in the current time slot or frame varying in the timescales of milliseconds or
seconds \cite{EEECLY,Amir2013Energy,Changyang2015Tcom,yang2018energy,xu2018energy}. Different from these policies, PRA leverages
future information in a minutes-long window \cite{Hatem2014EE,Ramy2017TWC,Ramy2018GCT}. By predicting trajectory \cite{LSTM_highway,ICSP18} and constructing radio map \cite{Radiomap,VTC18radiomap}, the future average channel gain (also called large-scale channel gain interchangeably in the sequel) in each frame can be predicted \cite{ICSP18}. By using the historical record of the modulation and coding scheme for a mobile user, its average data rate in each frame of a 30-seconds time window is  predicted in \cite{NB2018}. Though predicting information in such horizon is possible with machine learning, the prediction itself incurs extra costs for training and data gathering \cite{LSTM_highway,ICSP18,Radiomap,VTC18radiomap}. To achieve the gain of PRA at affordable costs, it is critical to study what information needs to be predicted and how to exploit different information effectively.


Maximizing the EE of a network should not compromise the quality-of-service (QoS) of users. Future cellular networks need to support diverse services with different QoS provision \cite{A2014Scenarios}. One type is real-time (RT) services such as video conference and voice over IP that require stringent QoS \cite{A2014Scenarios}. For this type of services, a data packet becomes useless once its required delay is violated. Hence, the QoS is characterized by the
\emph{statistical QoS requirement}, defined as a delay bound and
a delay bound violation probability, whose values depend on specific service \cite{3GPPQoS}.
The other type is non-real-time (NRT) services such as file downloading and video-on-demand (VoD). For VoD services,
the video quality and playback interruption are key metrics for user experience \cite{Juluri2015VoD}.
To meet the demands of different services efficiently, softwarization techniques such as network function virtualization (NFV) and software-defined networking (SDN) are proposed for the fifth generation (5G) networks \cite{View20175GPPP,bera2017software}. SDN manages radio resources and traffic flows in a centralized manner with a global view of the network state \cite{bera2017software}. NFV is a viable way to provide a network
slice tailored to each service \cite{ksentini2017toward}. In fact, with the global view of future average channel gains, the network performance can be improved by jointly optimizing predictive resource allocation for different types of services subject to the QoS of each user. However, existing works in the area of NFV/SDN focus on how to meet the demands of each kind of services rather than ensure the QoS of each user, and do not investigate how to harness the predictable trajectories of mobile users.

\vspace{-3.5mm}\subsection{Related Works}
Predictive resource allocation has been optimized separately for RT and NRT services.

For users requesting RT services, PRA is usually designed for improving admission
level QoS via mobility management with cell-level mobility prediction  \cite{Seng2006,Apollinaire2015Mobility,bui2016anticipatory,CGZ2018ICC}. By predicting the future handoff time and the BS that a RT user will access to, the bandwidth at the next BS was reserved for the user \cite{Seng2006}, and a call admission control scheme was proposed in \cite{Apollinaire2015Mobility}. By predicting the next several cells a RT user will enter, the delay caused by handoff and signaling is reduced significantly \cite{CGZ2018ICC}.

For users requesting NRT services, PRA is usually designed for boosting network performance such as EE or QoS of mobile NRT users with fine-grained prediction \cite{Hatem2014MSWiM,Hatem2014EE,DA2016ICC,Suhail2017ICC,Ramy2017TWC,Ramy2018GCT,Robert2014INFOCOM,Yao2016Planning}. Most existing works of PRA consider VoD service. With known future instantaneous channel gains, the trade-off between the required resources and the stalling time was investigated in \cite{DA2016ICC}. With known future average data rates in the frames of a prediction window, the number of time slots in each frame was optimized in \cite{Hatem2014EE} to save energy for ensuring the QoS of each VoD user.  In \cite{Hatem2014MSWiM}, a practical two-timescale PRA was proposed for LTE systems. In the first timescale, the number of time slots is optimized based on the rate prediction at the start of the prediction window, while in the second timescale the subcarriers are allocated in each time slot based on the instantaneous channel gains. Considering that future data rates cannot be predicted without errors, a robust PRA policy was optimized in \cite{Ramy2017TWC} by assuming bounded prediction errors. Further considering that the time resource occupied by RT services is uncertain due to the random request arrival, a robust PRA for VoD service was optimized in \cite{Ramy2018GCT}. A common assumption in \cite{Hatem2014EE,Hatem2014MSWiM,Ramy2017TWC,Ramy2018GCT} is that the future data rate of each user is predictable. However, the data rate of a wireless link depends on the resource allocation among users, i.e., the rate prediction is coupled with predictive resource allocation. This suggests that PRA with rate prediction is non-optimal. There also exist a few works of PRA considering the service of
file downloading \cite{Robert2014INFOCOM,Yao2016Planning}. By using future average channel gains, a proportional fair scheduling policy  was proposed in \cite{Robert2014INFOCOM}. With both future average channel gains of NRT users and average arrival rate of RT traffic, an energy-saving PRA policy was proposed  in \cite{Yao2016Planning}, where radio resources are reserved for RT services.

\vspace{-4mm}
\subsection{Motivations and Contributions}
All previous PRA policies are only optimized for a single kind of services. All policies are either optimized in one timescale or separately designed in two timescales. All existing policies are designed based on the fine-grained information (say trajectory or rate in each second) of every mobile user. While technically viable, predicting fine-grained information incurs high costs. For example, to predict fine-grained average channel gains, one needs to predict a fine-grained trajectory for every user. This requires a large number of training samples and high computational complexity for the off-line training\cite{ICSP18}. Besides, one needs to establish a fine-grained radio map for the network, where the average channel gains between each location and surrounding BSs need to be measured and stored, say by expensive drive tests \cite{VTC18radiomap}.

PRA policies can be optimized toward different objectives, which need very different techniques to find the optimal solutions. In this paper, we consider a not-fully-loaded network. While throughput-maximal PRA can boost the maximal number of users/requests that the network is able to support, EE-maximal PRA can save resources when the traffic load is not heavy, which is often the case in real-world cellular networks \cite{NB2018}.

Despite that prior works have demonstrated the potential of PRA, the following questions, which are important before
PRA is put into practice use, remain open: 1) To maximize the EE of a network, do we need to jointly optimize PRA for different types of services over multiple timescales? 2) Which kinds of future channel information are needed to maximize EE? 3) Is it possible to approach the maximal EE with coarse-grained future information? We strive to answer these questions in
this paper. Since the majority of data traffic is from mobile videos, we take VoD as an example of NRT services.  Our major contributions are summarized as follows:
\begin{itemize}
\item To show the potential of the joint optimization, we propose a framework to joint optimize PRA that maximizes the EE of network subject to the QoS requirements for both VoD and RT users by using two-timescale channel information. Finding the optimal solution is challenging, because optimizing the policies in two timescales turns out a functional extreme problem, which can not be solved by directly using convex optimization tools. To provide a baseline of comparison for the heuristic policy with coarse-grained prediction, we assume that the fine-grained future average channel gains for both types of users are perfectly known as in the existing works. Simulation results show that jointly optimizing PRA for both types of services and for two timescales can improve EE significantly.
\item To show which kind of future information is necessary to maximize EE, we analyze the degenerated optimization problem for the system serving only RT or VoD users. We find that predicting the average channel gains in the prediction window is helpful, but further predicting instantaneous channel gains in future time slots cannot improve EE.
\item To illustrate that PRA can achieve high EE even with coarse-grained prediction, we propose a heuristic policy, inspired by the structure of the optimization problem. This policy only needs the median of future average channel gains of VoD users. Surprisingly, the heuristic policy outperforms the optimal policy when the prediction errors are large, thanks to the fact that a median is insensitive to errors.
\end{itemize}


\vspace{-4mm}\section{System Model and QoS requirements}
Consider the scenario that multiple mobile users travel across the cells of an orthogonal frequency division multiple access (OFDMA) network. A user either requests for VoD or requests for RT service. For notational simplicity, we first consider a single cell scenario in this section and then extend to the multi-cell scenario at the end of the next section.

\vspace{-4mm}\subsection{Transmission and Channel Models}
Consider frequency-selective block fading channel. Time is discretized to frames each with duration $\Delta T$ and time slots each with duration $\tau$. The durations are defined according to the variation of large-scale channel gain and small-scale
channel gain due to user mobility, respectively.  The large-scale channel gains are predictable within a prediction window, with the predicted trajectories and a measured radio map \cite{Hatem2014EE}. The small-scale channel gains  (called instantaneous channel gains interchangeably in this work, also called CSI in literature) are predictable \cite{VTC18CSIprediction} within the channel coherence time (i.e., within $\tau$). For simplicity, we assume that: (1) the large-scale channel gain remains constant within
each frame and may vary among frames, and (2) the small-scale channel gain remains constant within
each time slot and is independent and identically distributed (i.i.d.) among time slots  in each frame and subcarriers.

Each frame includes $N_{\rm S}$ time slots, i.e., $\Delta T = N_S \tau$. A prediction window includes $N_L$ successive frames. At the beginning of a prediction window, the average channel gains in future frames within the window of both types of users are assumed known at the BS. However, CSI  is only known at the BS and the user at the beginning of each time slot.

There are $M_D+M_R$ users that access to the BS at the beginning of a prediction window,
where $M_D$ and $M_R$ are the numbers of users requesting VoD and RT services, respectively.
For the $m$th user, ${\alpha^m_i}$ is the average channel gain in the $i$th frame, and ${g^m_{ijk}}$ is the CSI on the $k$th subcarrier in the $j$th time slot of the $i$th frame.

The achievable instantaneous data rate for the $m$th user can be expressed as follows,  
\begin{align}
{s^{m}_{ij}} = B \sum\limits_{k = 1}^{{K^{m}_i}} {{{\log }_2}\left( {1 + \frac{{{\alpha^{m}_i}}}{{\phi \sigma _0^2}}{p^{m}_{ijk}}{g^{m}_{ijk}}} \right)}\quad \text{bits/s}, \label{eq:st}
\end{align}
where $B$ is the subcarrier spacing, $p^{m}_{ijk}$ is the transmit power allocated to the $m$th user  on the $k$th subcarrier in the $j$th time slot of the $i$th frame, $\phi>1$ captures the gap between capacity and achievable rate with practical modulation and coding schemes, $\sigma_0^2$ is the variance of the additive Gaussian noise, and $K^m_i$ is the number of subcarriers assigned to the $m$th user in the $i$th frame.

\vspace{-3mm}\subsection{QoS Requirement for VoD Service}
\begin{figure}[htbp]  
\vspace{-0.4cm}
\centering
\subfigure[{Queueing model for the $m$th VoD user.}]{
\label{fig:MUqueue} 
\includegraphics[width=0.48\textwidth]{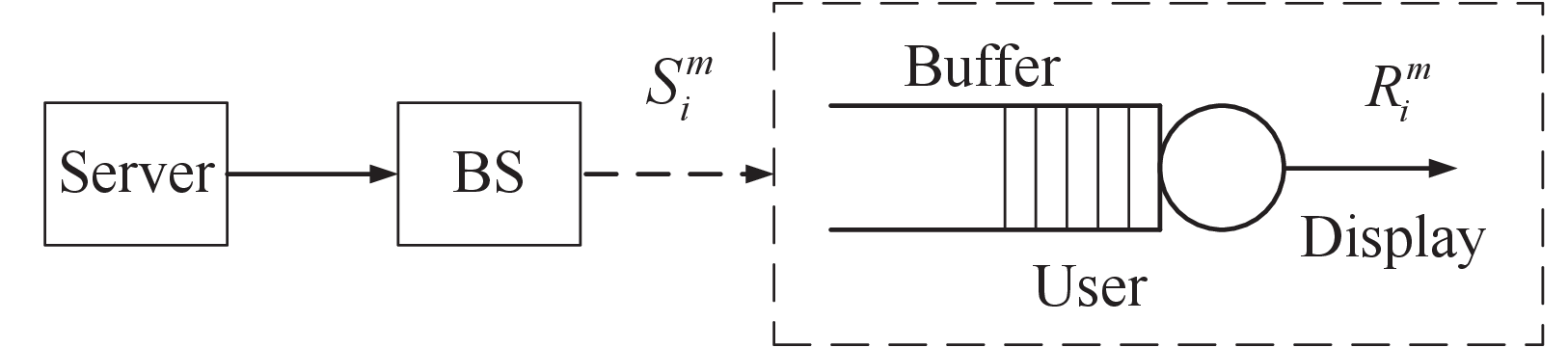}}
\subfigure[{Queueing model for the $m$th RT user.}]{
\label{fig:BSqueue} 
\vspace{-0.2cm}
\includegraphics[width=0.42\textwidth]{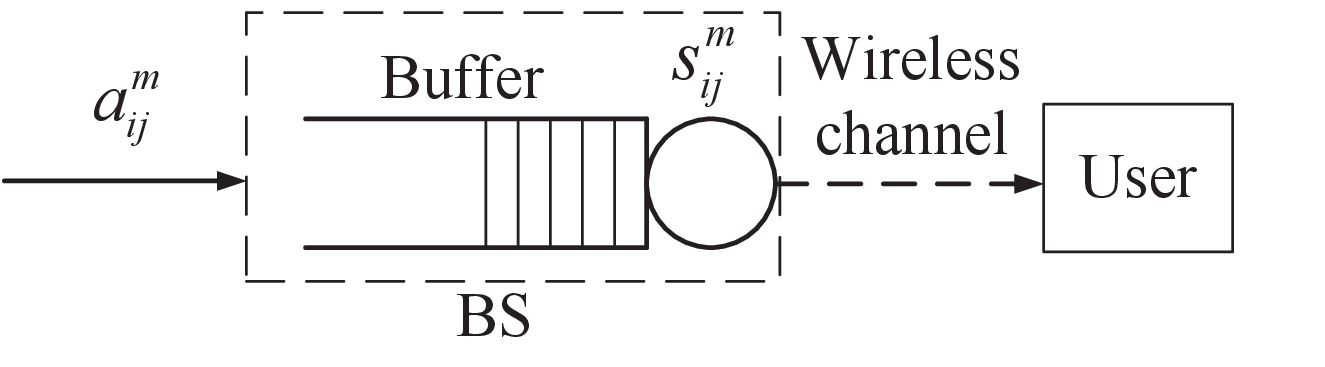}}
\caption{Queueing models for VoD and RT services.}
 \label{queue} 
\vspace{-0.4cm}
\end{figure}

Since the key factor that determines the experience of a user requesting a VoD service is playback interruption, we consider the queue in the buffer at each user. We assume that the video segments to be played within the prediction window are available at the BS as in \cite{Hatem2014EE,Ramy2017TWC,Apollinaire2016ICC}. The queueing model for VoD service is shown in Fig. \ref{fig:MUqueue}, where $R^m_i$ is the amount of data played at the $m$th user in the $i$th frame. The value of $R^m_i$ is given when a certain quality level of the video is chosen by the user (e.g., high definition video). The amount of data that can be transmitted to the $m$th user during the $i$th frame is given by ${S^m_i} = \tau\sum\limits_{j = 1}^{{N_S}} {{s^m_{ij}}}$.
%

Denote the duration of each video segment as $T_{seg}$. Without loss of generality, we set $T_{seg} = \Delta T$ for notational simplicity. Then, there are $N_L$ video segments in a prediction window. Assume that the buffer size is larger than the size of $N_L$ video segments. This is reasonable for smartphones since storage devices are cheap nowadays. The assumption will be removed in Section IV, where we design a heuristic policy that is aware of limited buffer size.

To avoid stalling during playback, each video segment should be delivered to {a VoD} user before the {segment} is played. Then, the QoS required by the VoD user can be reflected by the following constraint \cite{Hatem2014EE},
\begin{align}
{Q^m_0} + \sum\limits_{i = 1}^{l} {{{S^m_i}} }  \ge \sum\limits_{i = 1}^{l+1} { {{R^m_i}} } ,l = 1,...,{N_L}, m=1,...,M_D, \label{eq:play}
\end{align}
where ${Q^m_0} = {R^m_1} $ is the initial queue length and ${R^m_{N_L+1}}$ is the number of bits in the first video segment to be played in the next prediction window. Hence, no interruption occurs between the adjacent prediction windows.\footnote{At the beginning of the first prediction window, the user only needs to download the video segment played in the first frame. This will not lead to long waiting time at the beginning of the service.} {Scalable video coding (SVC) is used to encode videos, i.e., each video segment is encoded into one base layer and multiple enhancement layers\cite{Patrick2012Video}. When the channel quality is not good such that the data rate cannot satisfy the requirement in \eqref{eq:play}, we can reduce the value of $R^m_i$ by not transmitting some enhancement layers. In this way, we can reduce the stalling probability at the cost of sacrificing the definition of the video.}

Since the number of time slots in each frame is large in practice, by channel coding among time slots, the time-average data rate in a frame can approach the ensemble-average data rate  \cite{Neely2015Adaptive}.  From \eqref{eq:st}, the average data rate for the $m$th user in the $i$th frame can be expressed as,
\begin{align}
{\bar{s}^m_{i}} = B \sum\limits_{k = 1}^{{K^{m}_i}} {\mathbb{E}}_h\left[{{{\log }_2}\left( {1 + \frac{{{\alpha^{m}_i}}}{{\phi \sigma _0^2}}{p^{m}_{ijk}}{g^{m}_{ijk}}} \right)}\right]\quad \text{bits/s}, \label{eq:avest}
\end{align}
where the average is taken over small-scale channel fading. Then, we have ${S^m_i} = {\Delta T}{\bar{s}^m_{i}} $, and the QoS constraint in \eqref{eq:play} can be equivalently written as

\begin{align}
\sum\limits_{i = 1}^l { \bar{s}^m_{i} }  \ge \frac{1}{{\Delta T}}\sum\limits_{i = 2}^{l + 1} {{R^m_i}}, l=1,...,N_{L}, m=1,...,M_D. \label{eq:play2}
\end{align}

\vspace{-2mm}\begin{rem}
\emph{Other NRT services such as file downloading, whose user demand can be characterized as to transmit a file with size $\tilde{R}^m$ in $N_L$ frames, can also be included in our framework. Its QoS requirement can be expressed as $\sum\limits_{i = 1}^{N_L} { \bar{s}^m_{i} }  \ge \tilde{R}^m$, which is similar to \eqref{eq:play2}. }
\end{rem}

\vspace{-5mm}\subsection{QoS Requirement for Real-time Service}
{The} queueing model for the $m$th user requesting RT service is shown in Fig. \ref{fig:BSqueue}, where $a^m_{ij}$ represents the data arrival rate in the $j$th time slot of the $i$th frame. If the queueing delay in the $m$th queue exceeds a delay bound $D^m_{\max}$ with a delay violation probability less than $\varepsilon _D^m$, then the QoS requirement of the $m$th RT user can be satisfied \cite{Tang2007QoS,RALIU,Amir2013Energy,EEECLY}.


Effective bandwidth and effective capacity are widely applied tools in designing resource allocation with such statistical QoS requirement\cite{EB,EC}.\footnote{The term ``effective bandwidth" is not the spectrum resource in radio access networks. According to the definition in \cite{EB}, it is the minimal constant service rate that is required to ensure the QoS of a RT user with random arrived packets.} For uncorrelated random arrival process and service process, $\{a^m_{ij}\}$ and $\{s_{ij}^m\}$, the effective bandwidth and effective capacity can be expressed as $E_{{B}}^m\left( {\theta^m} \right) = \frac{1}{{\theta^m \tau }}\ln {\mathbb{E}}\left[ {{\exp\left({\theta^m \tau {a_{ij}^m} }\right)}} \right]$  (bits/s) {and}
\begin{align}
E_{{C_i}}^m\left( {{\theta^m}} \right) =  - \frac{1}{{\theta^m \tau }}\ln {\mathbb{E}}_{{g^m_{ijk}}}\left[ {{\exp\left({ - \theta^m \tau s_{ij}^m}\right)}} | \alpha_i^m \right]\quad \text{(bits/s)} \label{eq:EC},
\end{align}
respectively \cite{EB,EC}, where where $\theta^m$ is the \emph{QoS exponent}. The required QoS exponent $\theta^m$ to guarantee $(D^m_{\max},\varepsilon _D^m)$ can be obtained from \cite{Amir2013Energy}, i.e.,
\begin{align}
\Pr\{D_\infty^m > D^m_{\max}\} \approx \exp\left[-\theta^m E_{B}^m(\theta^m)D^m_{\max} \right] = \varepsilon _D^m \label{eq:reqtheta},
\end{align}
where $D^m_\infty$ is the steady state delay for the $m$th user. To ensure the QoS of the $m$th RT user over wireless channels, the following constraint should be satisfied \cite{RALIU}
\begin{align}
E_{{C_i}}^m\left( {{\theta^m}} \right) \geq E_{{B}}^m\left( {\theta^m} \right), m = M_D+1,...,M_D+M_R, i=1,...,N_L. \label{eq:EBEC}
\end{align}

\vspace{-6mm}\subsection{Power Consumption Model and EE Definition}
The total energy consumed by transmit power and circuit power at the BS for serving $M_D+M_R$ users in the prediction window (i.e., in $N_L$ frames) can be modeled as \cite{Claude2012Flexible}
\begin{align}
\sum\limits_{i = 1}^{N_L}{E_{i}} = \sum\limits_{i = 1}^{N_L}{\left(\frac{1}{\rho }\sum\limits_{m = 1}^{{M_D} + {M_R}} {\sum\limits_{j = 1}^{{N_S}} {\sum\limits_{k = 1}^{{K^{m}_i}} {\tau p^{m}_{ijk}} } }  + \Delta T{P_c}\sum\limits_{m = 1}^{{M_D} + {M_R}} {K^{m}_i}  + \Delta T{P_0}\right)},\label{eq:powermodel}
\end{align}
where $E_{i}$ is the energy consumption in the $i$th frame, $\rho \in (0,1]$ is the power amplifier efficiency, $P_{c}$ is the circuit power consumed for baseband processing such as channel estimation on each subcarrier, and $P_{0}$ is the fixed circuit power consumption for the BS.

According to the \emph{bits per Joule} metric in \cite{earth2010}, EE of a system is the ratio of the amount of  data transmitted to the energy consumed during a certain period. For PRA, the period is the duration of the prediction window. However, since only the average channel gains are available at the beginning of the prediction window, both the amount of data to be transmitted and the energy to be consumed in the upcoming $N_L$ frames are random variables, which depend on the instantaneous channel gains. As a result, we cannot optimize PRA to maximize the EE metric in \cite{earth2010}. Since the number of time slots in each frame is large, i.e., $N_S$ is large, maximizing the above EE metric is equivalent to maximizing the ratio of the average amount of  transmitted data to the average energy consumption, where the average is taken over the small-scale channel gains. Hence, we define the EE as follows,

\begin{align}
\eta \triangleq  {{\left[ {\mathbb{E}}_h\left(\sum\limits_{m = 1}^{{M_D} }{\sum\limits_{i = 1}^{{N_L}} \tau{\sum\limits_{j = 1}^{{N_S}} {{s^m_{ij}}}}}\right) + {\mathbb{E}}_h\left(\sum\limits_{m = M_D+1}^{{M_D+M_R} }{\sum\limits_{i = 1}^{{N_L}} \tau{\sum\limits_{j = 1}^{{N_S}} {{b^m_{ij}}}}} \right) \right]} \mathord{\left/
 {\vphantom {{\left[ {\mathbb{E}}_h\left(\sum\limits_{m = 1}^{{M_D} }{\sum\limits_{i = 1}^{{N_L}} \tau{\sum\limits_{j = 1}^{{N_S}} {{s^m_{ij}}}}}\right) + {\mathbb{E}}_h\left(\sum\limits_{m = M_D+1}^{{M_D+M_R} }{\sum\limits_{i = 1}^{{N_L}} \tau{\sum\limits_{j = 1}^{{N_S}} {{b^m_{ij}}}}} \right) \right]} {\left[ {\mathbb{E}}_h\left({\sum\limits_{i = 1}^{{N_L}} {{E_{i}}} }\right) \right]}}} \right.
 \kern-\nulldelimiterspace} {\left[ {\mathbb{E}}_h\left({\sum\limits_{i = 1}^{{N_L}} {{E_{i}}} }\right) \right]}}.\label{eq:EE}
\end{align}
For VoD services, the amount of data  transmitted equals the amount of data that needs to transmit. Thus, ${\mathbb{E}}_h\left(\sum\limits_{m = 1}^{{M_D} }{\sum\limits_{i = 1}^{{N_L}} \tau{\sum\limits_{j = 1}^{{N_S}} {{s^m_{ij}}}}}\right) = \sum\limits_{m = 1}^{{M_D} }{\sum\limits_{i = 1}^{N_L} {\Delta T \bar{s}^m_{i} } } = \sum\limits_{m = 1}^{{M_D} }{\sum\limits_{i = 2}^{{N_L}+1} {{R^m_i}}}$, which is determined at the beginning of the prediction window by the requested video level and network status.
For RT services, when the queues are in steady states, the average departure rates equal to the average arrival rates \cite{she2016energy}. Thus, ${\mathbb{E}}_h\left(\sum\limits_{m = M_D+1}^{{M_D+M_R} }{\sum\limits_{i = 1}^{{N_L}} \tau{\sum\limits_{j = 1}^{{N_S}} {{b^m_{ij}}}}} \right) = {\mathbb{E}_h}\left(\sum\limits_{m = M_D+1}^{{M_D+M_R} }{\sum\limits_{i = 1}^{{N_L}} \tau{\sum\limits_{j = 1}^{{N_S}} {{a^m_{ij}}}}} \right)$, which is determined by the arrival processes. Therefore, the numerator of \eqref{eq:EE} does not depend on the resource allocation. Further considering that the last term in \eqref{eq:powermodel} is a constant, maximizing EE is equivalent to minimizing the following average energy consumption,
\begin{align}
\frac{1}{\rho } {\mathbb{E}}_h\left(\sum\limits_{m = 1}^{{M_D} + {M_R}} {\sum\limits_{i = 1}^{N_L }{\sum\limits_{j = 1}^{{N_S}} {\sum\limits_{k = 1}^{{K^{m}_i}} {\tau p^{m}_{ijk}} } }}\right)  + \Delta T{P_c}\sum\limits_{m = 1}^{{M_D} + {M_R}} {\sum\limits_{i = 1}^{N_L }{K^{m}_i}}. \label{eq:objective}
\end{align}

\vspace{-3mm}
\section{Energy Efficient Predictive Resource Allocation}
In this section, we optimize predictive resource allocation under the assumption that the average channel gains in future frames of the window are perfectly known. We formulate a functional extreme problem and obtain the global optimal solution, referred to as \emph{ideal policy} for short. We first consider the single cell scenario and then extend to the multi-cell scenario.

\vspace{-4mm}\subsection{Problem Formulation}
At the beginning of a prediction window, we cannot optimize $p^m_{ijk}$ to minimize \eqref{eq:objective} since the instantaneous channel gains in future time slots are unknown. Yet we can optimize the average transmit power ${{\bar P}^m_i} \triangleq {\mathbb{E}}_h\left({{\sum\limits_{k = 1}^{{K^{m}_i}} { p^{m}_{ijk}} }}\right)$ and the number of subcarriers (i.e., bandwidth) $K^m_i$ assigned to the $m$th user in the $i$th frame, since the future average channel gains are known. We refer to $\{\bar{P}^m_i, K^m_i\}, m = 1,...,M_D+M_R, i=1,...,N_L$, as the \emph{resource allocation plan}. It determines the amount of resources assigned to the users in all frames of the prediction window.

At the beginning of each time slot, we can optimize $p_{ijk}^m$ according to the assigned resources in the corresponding frame $\{\bar{P}^m_i, K^m_i\}$, since the instantaneous channel gains $g_{ijk}^m$, $k=1,...,K^m_i$ are available at the BS. To gain useful insight, here we only consider power allocation, but not subcarrier allocation. We denote the \emph{power allocation policies} for the VoD users and the RT users as $p_{ijk}^m = f_D(\bar{P}^m_i, K^m_i, g_{ijk}^m), m = 1,...,M_D$ and $p_{ijk}^m = f_R(\bar{P}^m_i, K^m_i, g_{ijk}^m), m = M_D+1,...,M_D+M_R$, respectively, where $i = 1,...,N_L$, $j=1,...,N_S$ and $k = 1,...,K^m_i$. The forms of the functions $f_D(\cdot)$ and $f_R(\cdot)$ differ for different power allocation policies.

The optimization of resource allocation plan and power allocation policies are coupled. In what follows, we formulate the joint optimization problem for the two-timescale policy. We take Rayleigh fading as an example, but the methodology can be extended to the other channels.

Substituting the power allocation policy for
VoD service $p_{ijk}^m = f_D(\bar{P}^m_i, K^m_i, g_{ijk}^m)$ into \eqref{eq:avest}, the average service rate in the $i$th frame for Rayleigh fading can be expressed as follows,
\begin{align}
\bar s_i^m = K_i^m\int_0^\infty  {B{{\log }_2}\left[ {1 + \frac{{\alpha _i^m}}{{\phi \sigma _0^2}}{f_D}\left( {\bar P_i^m,K_i^m,g} \right)g} \right]{e^{ - g}}{\rm d}g}\label{eq:intst},
\end{align}
where $m = 1,...,M_D$, and $g$ is exponentially distributed with the mean of $1$.

Substituting the power allocation policy for
RT service $p_{ijk}^m = f_R(\bar{P}^m_i, K^m_i, g_{ijk}^m)$ into \eqref{eq:st}  and then into \eqref{eq:EC}, the effective capacity in the $i$th frame for Rayleigh fading can be obtained as
\begin{align}
E_{{C_i}}^m\left( {\theta^m} \right) =  - \frac{K_i^{m}}{{\theta^m\tau }}\ln \left\{ \int_0^\infty  {{{\left[ {1 + \frac{{\alpha _i^m}}{{\phi \sigma _0^2}}{f_R}\left( {\bar P_i^m,K_i^m,g} \right)g} \right]}^{ - {\beta ^m}}}{e^{ - g}}{\rm d}g}\right\} \quad \text{(bits/s)} ,\label{eq:intEC}
\end{align}
where $m = M_D+1,...,M_D+M_R$, and $\beta^m \triangleq \frac{\theta^m \tau B}{\ln 2}$.

Then, the optimal two-timescale policy that maximizes the EE with satisfied QoS requirement of each RT user and each VoD user can be obtained by solving the following problem,
\begin{align}
&\mathop {\mathop {\mathop {\min }\limits_{f_D(\cdot),f_R(\cdot), \bar P_i^m,K_i^m,} }\limits_{i = 1,...,{N_L},} }\limits_{m = 1,...,{M_D} + {M_R}} {E_{\rm ave}} \buildrel \Delta \over = \sum\limits_{m = 1}^{{M_D} + {M_R}} {\sum\limits_{i = 1}^{{N_L}} {\left( {\frac{1}{\rho }\bar P_i^m + {P_c}K_i^m} \right)} } , \label{eq:aimenergy}\\
\text{s.t.} & \quad \sum\limits_{i = 1}^l {K_i^m\int_0^\infty  {B{{\log }_2}\left[ {1 + \frac{{\alpha _i^m}}{{\phi \sigma _0^2}}{f_D}\left( {\bar P_i^m,K_i^m,g} \right)g} \right]{e^{ - g}}{\rm d}g} }  \ge \frac{1}{{\Delta T}}\sum\limits_{i = 2}^{l + 1} {R_i^m},\nonumber\\
& \quad m = 1,...,M_D, l = 1,..., N_L, \label{eq:Cst}\tag{\theequation a}\\
& \quad - \frac{{K_i^m}}{{{\theta ^m}\tau }}\ln \left\{ {\int_0^\infty  {{{\left[ {1 + \frac{{\alpha _i^m}}{{\phi \sigma _0^2}}{f_R}\left( {\bar P_i^m,K_i^m,g} \right)g} \right]}^{ - {\beta ^m}}}{e^{ - g}}{\rm d}g} } \right\} \ge E_B^m\left( {{\theta ^m}} \right), \nonumber\\
& \quad m = M_D+1,...,M_D+M_R, i = 1,...,N_L, \label{eq:CEC}\tag{\theequation b}\\
&\quad \sum\limits_{m = 1}^{{M_D} + {M_R}} {\bar P_i^m}  \le {P^{\max}_{\rm ave }}, i = 1,...,{N_L},\label{eq:maxpower}\tag{\theequation c}\\
&\quad \sum\limits_{m = 1}^{{M_D} + {M_R}} {K_i^m}  \le {K_{\max}}, i = 1,...,{N_L},\label{eq:maxK}\tag{\theequation d}\\
&\quad  \bar P_i^m \ge 0, K_i^m \ge 0, K_i^m \in \mathbb{Z}, m = 1,..., M_D+M_R,i = 1,...,{N_L}, \label{eq:positive}\tag{\theequation e}
\end{align}
where the objective function in \eqref{eq:aimenergy} is obtained by substituting ${\bar P^m_i} = {{\mathbb{E}_h}}\left( {{\sum\limits_{k = 1}^{{K^m_i}} {p_{ijk}^m} } } \right)$ into \eqref{eq:objective} and ignoring a constant $\Delta T = N_S \tau$, constraints in \eqref{eq:Cst} and \eqref{eq:CEC} are obtained by substituting \eqref{eq:intst} and \eqref{eq:intEC} into \eqref{eq:play2} and \eqref{eq:EBEC}, respectively, and constraints in \eqref{eq:maxpower} and \eqref{eq:maxK} ensure that the average transmit power and the number of subcarriers allocated to all the users do not exceed the maximal average transmit power $P^{\max}_{\rm ave }$ and the total number of subcarriers $K_{\max}$.
With constraint \eqref{eq:maxK}, we can always allocate each subcarrier only to one user at each time slot. {Due to the bandwidth and power constraints, the problem could be infeasible when the system is heavy loaded and the channels are not good. In this case, the system can drop some enhancement layers of the SVC for VoD service, and then the value of $R_i^m$ is reduced. To minimize the quality deterioration, $R_i^m$ should be optimized when the system is fully loaded as in the literature, e.g., \cite{Riiser2012Video,Xuan2015hotmobile}. In this work, we study how to improve EE when the system is not fully loaded. In this case, the problem is feasible.}

Finding the optimal solution of problem \eqref{eq:aimenergy} is non-trivial. On the one hand, the constraints in \eqref{eq:Cst} and \eqref{eq:CEC} depend on the forms of the functions $f_D(\cdot)$ and $f_R(\cdot)$. As a result, the optimal values of $E_{\rm ave}$ in \eqref{eq:aimenergy} is a function of power allocation policies. We denote the minimal energy consumption with given power allocation policies as $E_{\rm ave}^{*}\left(f_D,f_R\right)$. The optimal power allocation policies can be obtained by minimizing $E_{\rm ave}^{*}\left(f_D,f_R\right)$, and are denoted as $f_D^*(\cdot)$ and $f_R^*(\cdot)$. Finding the optimal form of functions {results in an optimization problem in the calculus of variations \cite{Gregory2018Constrained}, where the optimization variables are functions that can be regarded as vectors with infinite dimensions. Since convex optimization tools can only be used to solve finite dimensional optimization problems, they are not applicable here.} On the other hand, there are no closed-form expressions of constraints \eqref{eq:Cst} and \eqref{eq:CEC}. Although the achievable rate is concave in transmit power and bandwidth with equal power allocation \cite{EEECLY}, whether or not the convexity still holds with optimal power allocation policies is unknown.

To address the challenge of deriving the optimal two-timescale policy, we first find the functions of $f_D^*(\cdot)$ and $f_R^*(\cdot)$ that minimizes $E_{\rm ave}^{*}\left(f_D,f_R\right)$, by proving that two spectral-efficient power allocation policies are fortunately {also} able to maximize EE. Then, {we find} the optimal resource allocation planning from problem \eqref{eq:aimenergy} upon substituting to $f_D^*(\cdot)$ and $f_R^*(\cdot)$.

\vspace{-2mm}\begin{rem}
\emph{The terms inside the sum of the left-hand side of \eqref{eq:Cst} are the average rates in different frames  (i.e., $\bar s_i^m$ in \eqref{eq:intst}). In many existing works \cite{Hatem2014EE,Hatem2014MSWiM,Ramy2017TWC,Ramy2018GCT},
this average rate is assumed known by prediction.
However, it is clear from problem \eqref{eq:aimenergy} that the future average rate depends on $\{\bar{P}^m_i, K^m_i\}$ and $f_D(\cdot)$ even if the system only supports VoD services. This suggests that making the resource allocation plan based on average rate prediction is non-optimal.
}
\end{rem}

\vspace{-7mm}\subsection{Optimal Power Allocation Policies}
A policy that maximizes the average service rate $\bar{s}^m_i$ (or effective capacity $E^m_{C_i}(\theta^m)$) with given average transmit power $\bar{P}^m_i$ and number of subcarriers $K^m_i$ (i.e., bandwidth) can also minimize $\bar{P}^m_i$ with given $\bar{s}^m_i$ (or $E^m_{C_i}(\theta^m)$) and $K^m_i$ \cite{WirelessCom,Tang2007QoS}. Yet this does not mean that the policy is optimal to minimize the average energy consumption in \eqref{eq:aimenergy}, i.e., maximize the EE.

\subsubsection{Power allocation policy for VoD service} As shown in \cite{WirelessCom}, the policy that maximizes $\bar{s}^m_i$ with given $\bar{P}^m_i$ and $K^m_i$ is water-filling, which is
\begin{align}\label{eq:waterfill}
f_D^w\left(\frac{\bar{P}^m_i}{K^m_i},g\right) = \left\{ {\begin{array}{*{20}{c}}
   {\frac{\phi \sigma^2_0}{\alpha^m_i}\left(\frac{1}{{{{\nu}_i^{m}}}} - \frac{1}{{g}}\right),{g} \ge {{\nu}_i^{m}}},  \\
   {0,\quad\quad\quad{g} < {{\nu}_i^{m}}},  \\
\end{array}} \right.
\end{align}
where the water level ${\nu}_i^{m}$  can be obtained from
$\int_{{{\nu}_i^{m}}}^\infty  {\frac{\sigma^2_0}{\alpha^{m}_i}\left( {\frac{1}{{{{\nu}_i^{m}}}} - \frac{1}{g}} \right)e^{-g}{\rm d}g}  = \frac{\bar{P}^{m}_i}{K^{m}_i}$.

\subsubsection{Power allocation policy for RT service}
As shown in \cite{Tang2007QoS}, the policy that maximizes $E_{{C_i}}^m\left( {{\theta^m}} \right)$ with given $\bar{P}^{m}_i$ and $K^{m}_i$ also follows a water-filling structure, which is
\begin{align}\label{eq:waterfillRT}
f_R^w\left(\frac{\bar{P}^m_i}{K^m_i},g\right) = \left\{ {\begin{array}{*{20}{c}}
{\frac{{\phi \sigma _0^2}}{{{\alpha^m_i}}}\left[ {\frac{1}{{{{\left( {{\nu_i^{m}}} \right)}^{\frac{1}{{\beta^m  + 1}}}}{{ {g} }^{\frac{\beta^m }{{\beta^m  + 1}}}}}} - \frac{1}{{g}}} \right],g \ge {\nu_i^{m}},}\\
{0,\;\;\;{\kern 1pt} \;\;\;{\kern 1pt} \;\;\;{\kern 1pt} g < {\nu_i^{m}},}
\end{array}} \right.
\end{align}
where $m = M_D+1,...,M_D+M_R, i = 1,...,N_L$, $\beta^m = \frac{\theta^m \tau B}{\ln 2}$, and the water level ${\nu}_i^{m}$ over Rayleigh fading channel can be obtained from
\begin{align}\label{eq:cutoffRT}
\int_{{\nu_i^{m}}}^\infty  {\frac{{\phi \sigma _0^2}}{{\alpha _i^{m}}}\left[ {\frac{1}{{{{\left( {{\nu_i^{m}}} \right)}^{\frac{1}{{\beta^m  + 1}}}}{{ g }^{\frac{\beta^m }{{\beta^m  + 1}}}}}} - \frac{1}{{g}}} \right]{e^{ - g}}{\rm d}g}  = \frac{{\bar P_i^{m}}}{{K_i^{m}}}.
\end{align}
The water-level is time-varying and the instantaneous power allocated to each subcarrier depends on the instantaneous  channel gains on all the subcarriers assigned to the user.


\subsubsection{Optimality of the power allocation policies}
The following proposition (see proof in Appendix \ref{App:optimal}.) indicates that \eqref{eq:waterfill} is the optimal power allocation policy for VoD service and \eqref{eq:waterfillRT} is the optimal power allocation policy for RT service in terms of maximizing the EE. In other words,  $f^*_D(\bar{P}^m_i, K^m_i, g) = f_D^w\left(\frac{\bar{P}^m_i}{K^m_i},g\right)$ and $f^*_R(\bar{P}^m_i, K^m_i, g) = f_R^w\left(\frac{\bar{P}^m_i}{K^m_i},g\right)$.
\vspace{-2mm}\begin{prop}\label{P:optimal}
\emph{For ANY power allocation policies $f'_D\left({\bar{P}^m_i},{K^m_i},g\right)$ and $f'_R\left({\bar{P}^m_i},{K^m_i},g\right)$,}
\begin{align}
E^*_{\rm ave}\left(f_D^w, f_R^w\right) \leq E^*_{\rm ave}\left(f'_D, f'_R\right). \label{eq:optimalpolicy}
\end{align}
\end{prop}

\vspace{-8mm}\subsection{Optimal Resource Allocation Planning}
Substituting the optimal power allocation policies in \eqref{eq:waterfill} and \eqref{eq:waterfillRT} into \eqref{eq:Cst} and \eqref{eq:CEC}, the optimal resource allocation plan can be obtained from the following problem,
\begin{align}
&\mathop {\mathop {\mathop {\min }\limits_{\bar P_i^m,K_i^m,} }\limits_{{m = 1,...,{M_D} + {M_R}, {i = 1,...,{N_L}} }}} \sum\limits_{m = 1}^{{M_D} + {M_R}} {\sum\limits_{i = 1}^{{N_L}} {\left( {\frac{1}{\rho }\bar P_i^m + {P_c}K_i^m} \right)} } , \label{eq:aimpower}
\end{align}
\begin{align}
\text{s.t.} & \quad \sum\limits_{i = 1}^l {K_i^m{F_D}\left( {\frac{{\bar P_i^m}}{{K_i^m}}} \right)}  \ge \frac{1}{{\Delta T}}\sum\limits_{i = 2}^{l + 1} {R_i^m}, m = 1,...,M_D, l = 1,..., N_L, \label{eq:CVOD}\tag{\theequation a}\\
& \quad - \frac{{K_i^m}}{{{\theta ^m}\tau }}\ln \left[ {F_R}\left( {\frac{{\bar P_i^m}}{{K_i^m}}} \right)\right] \ge E_B^m\left( {{\theta ^m}} \right), m = M_D+1,...,M_D+M_R, i = 1,...,N_L, \label{eq:CRT}\tag{\theequation b}\\
&\quad \eqref{eq:maxpower}, \eqref{eq:maxK}\; \text{and} \; \eqref{eq:positive},\nonumber
\end{align}
where
\begin{align}
&{F_D}\left( {\frac{{\bar P_i^m}}{{K_i^m}}} \right) = \int_0^\infty  {B{{\log }_2}\left[ {1 + \frac{{\alpha _i^m}}{{\phi \sigma _0^2}}f_D^w\left( {\frac{{\bar P_i^m}}{{K_i^m}},g} \right)g} \right]{e^{ - g}}{\rm d}g},\; \label{eq:FVOD}\\
&{F_R}\left( {\frac{{\bar P_i^m}}{{K_i^m}}} \right) = \int_0^\infty  {{{\left[ {1 + \frac{{\alpha _i^m}}{{\phi \sigma _0^2}}f_R^w\left( {\frac{{\bar P_i^m}}{{K_i^m}},g} \right)g} \right]}^{ - {\beta ^m}}}{e^{ - g}}{\rm d}g}\label{eq:FRT}.
\end{align}

By relaxing the numbers of subcarriers as continuous variables, we can obtain the following property (See proof in Appendix \ref{App:PEC}).
\vspace{-1mm}\begin{pro}\label{P:EC}
\emph{The left-hand sides of \eqref{eq:CVOD} and \eqref{eq:CRT} are jointly concave in $\bar{P}^{m}_i$ and ${K_i^{m}}$.}
\end{pro}\vspace{-3mm}

%

The above property indicates that the feasible region of problem \eqref{eq:aimpower} is a convex set. Since the objective function in \eqref{eq:aimpower} is linear, problem \eqref{eq:aimpower} is convex programming, whose global optimal solution can be solved numerically by the interior-point method if it is feasible \cite{boyd}.

The \emph{ideal policy}, i.e., the optimal solution of problem \eqref{eq:aimenergy}, consists of making a plan and allocating power that operate in two timescales. The resource allocation plan for a user is made at the start of the prediction window with predicted average channel gains, which is optimized from problem \eqref{eq:aimpower}. The transmit power is allocated at the start of each time slot with estimated CSI, which is optimized from \eqref{eq:waterfill} and \eqref{eq:waterfillRT}.


\vspace{-3mm}\subsection{Impacts of Predicted Information on EE}
Predicting the instantaneous channel gains (i.e., CSI) and average channel gains of every user are  possible, but inevitably incurs cost for training \cite{ICSP18,VTC18CSIprediction}. In the sequel, we discuss which kinds of future information are necessary to maximize the EE.

\subsubsection{Predicted Information of VoD Users}
Here we consider a system without RT users, i.e., $M_R = 0$. If the future CSI is available at the BS at the beginning of each prediction window and $M_R = 0$, then minimizing \eqref{eq:objective} is equivalent to minimizing the following objective function,
\begin{align}
\frac{1}{\rho }\left(\sum\limits_{m = 1}^{{M_D}} {\sum\limits_{i = 1}^{N_L }{\sum\limits_{j = 1}^{{N_S}} {\sum\limits_{k = 1}^{{K^{m}_i}} {\tau p^{m}_{ijk}} } }}\right)  + \Delta T{P_c}\sum\limits_{m = 1}^{{M_D}} {\sum\limits_{i = 1}^{N_L }{K^{m}_i}}, \label{eq:objective2}
\end{align}
where the transmit powers on different subcarriers in the $i$th frame $\{p^{m}_{ijk},k=1,...,K^{m}_i\}$ depend on the instantaneous channel gains $\{g^{m}_{ijk},k=1,...,K^{m}_i\}$. Denote the total transmit power for the $m$th user in the $j$th time slot of the $i$th frame as $P^m_{ij} = {\sum\limits_{k = 1}^{{K^{m}_i}} {p^{m}_{ijk}} }$. Since the fast fading channels among the time slots in each frame are i.i.d., if the number of time slots in each frame is large,  then the time-average transmit power converges to the ensemble-average transmit power, i.e., $\frac{1}{N_S}\sum\limits_{j = 1}^{{N_S}}{P^m_{ij}} \to \bar{P}^m_i$ when $N_S \to \infty$. Further considering that $\Delta T = N_S \tau$, minimizing \eqref{eq:objective2} is equivalent to minimizing the following expression,
\begin{align}
\frac{1}{\rho }\left(\sum\limits_{m = 1}^{{M_D}} {\sum\limits_{i = 1}^{N_L }{ {{\bar{P}^{m}_{i}} } }}\right)  + {P_c}\sum\limits_{m = 1}^{{M_D}} {\sum\limits_{i = 1}^{N_L }{K^{m}_i}}, \label{eq:objective3}
\end{align}
which is the same as the objective function in \eqref{eq:aimenergy} when $M_R = 0$.

The ideal policy that minimizes the energy consumed for VoD users with future CSI can be obtained by minimizing \eqref{eq:objective3} under constraints \eqref{eq:Cst}, \eqref{eq:maxpower}, \eqref{eq:maxK} and \eqref{eq:positive}. Since the optimization problem is the same as problem \eqref{eq:aimenergy}, the optimal power allocation policies, the optimal average transmit power and numbers of subcarriers, and the minimal total energy consumptions obtained from the two problems are equal. This leads to the following observation.\vspace{2mm}\\
{\bf Observation 1}: Predicting the CSI of each VoD user in future time slots does not help improve the system EE, but predicting the average channel gains of the VoD user can improve EE.



\subsubsection{Predicted Information of RT Users}
Similarly, here we consider a system without VoD users, i.e., $M_D = 0$.
For RT service, $\tau \ll D^m_{\max} \ll \Delta T$, where $\tau$ and $\Delta T$ are the time slot and frame durations. At the beginning of each frame, when the average channel gain is available by estimation, the BS can assign the average transmit power and number of subcarriers to each RT user. The amount of resources assigned to the $m$th RT user in the $i$th frame can be obtained from the following problem,
\begin{align}
 \mathop {\min }\limits_{\bar P_i^m,K_i^m,}  & \sum\limits_{m = 1}^{ {M_R}} {{\left( {\frac{1}{\rho }\bar P_i^m + {P_c}K_i^m} \right)} } \label{eq:aimRT}\\
 \text{s.t.}& ~\eqref{eq:CEC}, \eqref{eq:maxpower}, \eqref{eq:maxK}\;\text{and}\; \eqref{eq:positive}.  \nonumber
\end{align}
From the expressions of the constraints, we can see that the amount of resources assigned in the $i$th frame does not depend on the amount of resources assigned in other frames. Therefore, problem \eqref{eq:aimenergy} can be decomposed into $N_L$ independent problems as problem \eqref{eq:aimRT}. Knowing the average channel gains (and hence CSI) in future frames cannot help improve the QoS (i.e., $D^m_{\max}$ and $\varepsilon _D^m$) or the EE \emph{of a system only with RT services}. This implies that making the resource allocation plan for RT users is unnecessary, and gives rise to another observation as follows.\vspace{1mm}\\
{\bf Observation 2}: Predicting the average channel gains and CSI of each RT user in the prediction window does not help improve the EE of a system only with RT service.

\vspace{-2mm}\begin{rem}\label{r:realtime}
\emph{VoD service is delay-tolerant. The QoS of a VoD user can be satisfied if the requested segment can be downloaded  to the user before playback. Hence, the BS can choose some frames in the prediction window with high average channel gains to transmit data in advance to save energy. By contrast, RT service is delay-sensitive. The QoS of a RT user in terms of delay (i.e., $D^m_{\max}$) is less than the frame duration $\Delta T$. Hence, to improve the EE under the QoS constraint, the BS can only adjust resources among the time slots within $D^m_{\max}$. This explains why the future average channel gains of each RT user cannot help improve the EE of a system only {with} RT users. Nevertheless, predicting the average channel gains of RT users helps improve the EE of a network with both VoD and RT services.}
\end{rem}

\vspace{-6mm}\subsection{Extension to Multicell Scenario}
Now we consider a scenario where the $M_D+M_R$ users are served by $N_B$ BSs.
The BSs are connected with a central processor (CP) and send the future average channel gains of all the users in a prediction window to the CP. To focus on the EE-maximal optimization for both services and in two timescales, we assume that the inter-cell interference can be treated as noise. A simple way to avoid strong interference is using orthogonal resources in adjacent cells, say by soft frequency reuse. This assumption is reasonable for the problem at hand, because we consider a non-fully-loaded network. The problem to optimize PRA with strong interference is nontrivial, as demonstrated in \cite{guo2018interference}, where a PRA is designed for file downloading in heterogeneous networks.

It is not hard to show that Proposition \ref{P:optimal} can be extended into the multi-cell scenario, and hence the power allocation policies in \eqref{eq:waterfill} and \eqref{eq:waterfillRT} are optimal for VoD service and RT service, respectively. Denote $\bar{P}^{mn}_i$ and $K^{mn}_i$ as the average transmit power and the number of subcarriers assigned to the $m$th user in the $i$th frame by its accessed BS (i.e., the $n$th BS). Denote ${\mathcal{M}}_i^n$ as the set of indices of the users that are served by the $n$th BS in the $i$th frame. The difference between single-cell and multi-cell scenarios lies in the constraints on the average transmit power and the total number of subcarriers. Specifically, the amount of resources assigned to the users that access to the same BS in the multi-cell scenario should satisfy the following constraints
\begin{align}
\sum\limits_{m \in {\mathcal{M}}_i^n} {\bar P_i^{mn}}  \le P^{\max}_{\rm ave }, \sum\limits_{m \in {\mathcal{M}}_i^n} {K_i^{mn}}  \le K_{\max }, n=1,...,N_B, i = 1,...,N_L. \label{eq:MUPmax}
\end{align}
The user association $\{{\mathcal{M}}_i^n, n=1,...,N_B, i = 1,...,N_L\}$ and resource allocation plan can be jointly optimized, but the resulting problem is a mixed integer optimization problem, which is much more challenging than the problem for the single-cell scenario. To save energy, it is reasonable to assume  that each user is accessed to the BS with the highest large-scale channel gain. Then, $\{{\mathcal{M}}_i^n, n=1,...,N_B, i = 1,...,N_L\}$ are known by the CP at the beginning of each prediction window with the predicted trajectories of mobile users. Similar to problem \eqref{eq:aimpower}, the optimal resource allocation plan is also convex programming.


\vspace{-3mm}\section{A Heuristic Joint Resource Allocation Policy with Low Costs}
In this section, we propose a heuristic policy that can perform close to the optimal solution with coarsely predicted knowledge. The queue states in the buffers of VoD users are also taken into account. To be more specific, each VoD user sends two bits information to the BS to indicate whether the buffer will overflow and whether playback interruption will occur.\footnote{In practice, the user can send a request for stopping transmission if the buffer will overflow. If the last video segment in the buffer is played in the current frame, the user can send a transmission request to avoid interruption in the next frame.}

This policy is inspired by the structure of problem  \eqref{eq:aimpower}, which suggests that the amount of resources assigned for a RT user  in each frame is independent of the resources assigned in other frames if only RT users exist. This implies that predictive resource allocation for RT users cannot improve EE. In other words, we only need to design the resource allocation plan for VoD users. If we can also decompose the resource allocation planning problem for VoD users into $N_L$ independent problems, then many existing low-complexity algorithms can be applied to assign resources in each frame for both types of users.

To decouple the resource allocation planning problem, we come back to the basic idea of PRA that only serves VoD users: transmit more data to a user when the user undergoes good average channels \cite{Hatem2014EE}. Such an idea can be translated as: which frame is with good channel to boost the
EE and how much data should be transmitted to satisfy the QoS.

To increase EE, we find a ``ruler'' to judge whether the average channel gain in a frame is high or low. According to the results in \cite{ICCC2015} obtained from optimizing for a single VoD user, the ideal policy transmits data only when the average channel gains exceed a certain threshold, which occurs over around half of the frames in the prediction window. This inspires us to use the median of the average channel gains in the window, denoted as $\alpha^m_{\rm med}$, as the threshold. Then, at the beginning of the window, the CP only needs to predict the median for each VoD user.

To avoid stalling and buffer overflow for the VoD users with limited buffer sizes, we should consider the queueing status of each VoD user and control the number of segments transmitted in each frame. The number depends on the traffic load of the network, the buffer size and channel condition of each VoD user. Since we use the median as the threshold, in average the BS transmits data to a VoD user in 50\% time slots during streaming. Then, it is reasonable to transmit several segments  (we consider two segments for illustration in the sequel, but more segments can be transmitted) to a user with good channel in a frame, if there is still room in the buffer. With given average power and bandwidth in each frame, the transmission procedure for each VoD user (say the $m$th user) of the heuristic policy is as follows.

At the beginning of the $i$th frame, the average channel gain of the $m$th user, $\alpha^m_{i}$, can be estimated at its associated BS. Denote $\tilde{i}$ as the index of last video segment that has been transmitted before the $i$th frame. Then, the indices of segments to be transmitted are $\{\tilde{i}+1,...\}$.

If $ \alpha^m_{i} < \alpha^m_{\rm med}$, then the $m$th user is in bad channel condition. No data will be transmitted in the $i$th frame if the video segment to be played in the $i+1$th frame has been transmitted before the $i$th frame. Otherwise, one segment will be transmitted in the $i$th frame. Thus, the required average service rate can be expressed as follows,
\begin{align}\label{eq:sbad}
\hat s_i^m = \left\{ {\begin{array}{*{20}{l}}
{0,\quad\quad\;\;\;\,{\rm{if}\;}\tilde i \ge i + 1},\\
{\frac{1}{{\Delta T}}R_{i + 1}^m,\;{\rm otherwise}.}
\end{array}} \right.
\end{align}
If $ \alpha^m_{i} \geq \alpha^m_{\rm med}$, then the $m$th user is in good channel condition. Two segments will be transmitted in the $i$th frame if the buffer has enough space for two segments. One segment will be transmitted if the buffer only has space for one more segment. If there is no enough space, no data will be transmitted. The required average service rate is given by
\begin{align}\label{eq:sgood}
\hat s_i^m = \left\{ {\begin{array}{*{20}{l}}
{\frac{1}{{\Delta T}}(R_{\tilde i + 1}^m + R_{\tilde i + 2}^m),{\rm{if \;}}Q_i^m + R_{\tilde i + 1}^m + R_{\tilde i + 2}^m - R_i^m \le {Q_{\max }},}\\
0,\quad\quad\quad\quad\quad\quad\;\;\,{{\rm{if \;}}Q_i^m + R_{\tilde i + 1}^m - R_i^m > {Q_{\max }},}\\
\frac{1}{{\Delta T}}R_{\tilde i + 1}^m,\quad\quad\quad\quad{\rm{otherwise}},
\end{array}} \right.
\end{align}
where $Q^m_i$ and  $Q_{\max}$ are the queue length of the $m$th user at the beginning of the $i$th frame and the maximal buffer size, respectively.

The average power and bandwidth assigned to each VoD and RT user in the $i$th frame can be jointly optimized from the following problem,
\begin{align}
&\mathop {\mathop {\mathop {\min }\limits_{\bar P_i^m,K_i^m,} } }\limits_{m = 1,...,{M_D} + {M_R}} \sum\limits_{m = 1}^{{M_D} + {M_R}} { {\left( {\frac{1}{\rho }\bar P_i^m + {P_c}K_i^m} \right)} } , \label{eq:Heuristic}\\
\text{s.t.} & \quad  {K_i^m{F_D}\left( {\frac{{\bar P_i^m}}{{K_i^m}}} \right)}  \ge \hat{s}^m_i, m = 1,...,M_D, \label{eq:HCVOD}\tag{\theequation a}\\
& \quad - \frac{{K_i^m}}{{{\theta ^m}\tau }}\ln \left[ {F_R}\left( {\frac{{\bar P_i^m}}{{K_i^m}}} \right)\right] \ge E_B^m\left( {{\theta ^m}} \right), m = M_D+1,...,M_D+M_R, \label{eq:HCRT}\tag{\theequation b}\\
&\quad \eqref{eq:maxpower}, \eqref{eq:maxK}\; \text{and} \; \eqref{eq:positive},\nonumber\vspace{-3mm}
\end{align}
where the optimal power allocation policies in \eqref{eq:waterfill} and \eqref{eq:waterfillRT} are adopted.

Compared with problem \eqref{eq:aimpower}, the only difference lies in the QoS constraints of VoD users on the average service rate in \eqref{eq:CVOD} and \eqref{eq:HCVOD}. In problem \eqref{eq:aimpower}, $\bar{s}_i^m$ is optimized implicitly through optimizing $K_i^m$ and $\bar P_i^m$ according to all average channel gains in the prediction window. In problem \eqref{eq:Heuristic} the value of $\hat{s}_i^m$ is only determined by the average channel gain in the $i$th frame and the threshold. According to the way we obtain $\hat{s}_i^m$ (i.e., \eqref{eq:sbad} and \eqref{eq:sgood}) in the heuristic policy, the average rate $\bar{s}_i^m$ will satisfy constraint \eqref{eq:CVOD} if constraint \eqref{eq:HCVOD} is satisfied. Thus, a feasible solution of problem \eqref{eq:aimpower} can be obtained with the heuristic policy.

Because the average service rate constraint in \eqref{eq:HCVOD} is a special case of the effective capacity constraint in \eqref{eq:HCRT} with $\theta^m \to 0$ \cite{EC}, many existing algorithms can be applied to find the solution of problem \eqref{eq:Heuristic} \cite{EEECLY}. This indicates that except the cost in predicting the ``ruler'' for each VoD user (i.e., $\alpha^m_{\rm med}$) at the CP, the heuristic policy needs the same complexity as existing non-predictive counterparts. After obtaining average power and bandwidth assigned to each VoD and RT user in each frame, the optimal power allocation policies in \eqref{eq:waterfill} and \eqref{eq:waterfillRT} can be used at each time slot. The heuristic policy is summarized in Table \ref{T:Heuristic}.

\begin{rem}\label{R:4}
\emph{With the heuristic policy, future information is required only when we determine $\hat{s}^m_i$ by the ``ruler'', which is the median of the future average channel gains in the prediction window of the $m$th VoD user, $\alpha^m_{\rm med}$.
{ To predict $\alpha^m_{\rm med}$, we can design a fully connected neural network (NN), which input is the historical average channel gains of VoD users  denoted as ${\bm x}$ and the output is the channel median denoted as $y$. The NN is trained with the training set $\{{\bm x}^{(n)}, {y}^{(n)}\}_{n=1}^N$ to minimize the cost function
$J({\bm {W,b}}) = \frac{1}{N} \sum_{n=1}^N |\hat{{y}}^{(n)} - {y}^{(n)}|^2 + \frac{\nu}{2} \|{\bm W}\|_{\mathrm F}^2$,
where ${\bm W}$ and ${\bm b}$ are the weight matrix between layers of the NN and the bias of neurons, ${{y}}^{(n)}$ is the label, $\hat{{y}}^{(n)}$ is the output with input ${\bm x}^{(n)}$, and the regularization term is added to reduce overfitting. By contrast, to predict the fine-grained average channel gain in each frame of the prediction window (say $\alpha_i^m$ ), recurrent neural networks need to be used for predicting fine-grained trajectory \cite{ICSP18}.
For training and testing the prediction, we consider mobile users with trajectories in \cite{ICSP18}, the only difference is that the length of each road segment is 1 km and the random stoping time at the red lights is 1 $\sim$ 30 s here to reduce training time. For a one minute-long prediction window, after tuning the hyper-parameters of our NN and the long short term memory model proposed in \cite{ICSP18}, the results are as follows. When the users move along a road with minimal distance from BSs as 200 m, the average relative error (i.e., prediction errors normalized by the true values in the test set divided by the number of test samples) of $\hat \alpha^m_{\rm med}$ is 42\% and that of $\hat \alpha_i^m$ is 170\%. The EE loss caused by prediction errors for the heuristic and optimal policies are 5.4\% and 65\%, respectively. 2000 and 56000 training samples are required for predicting $\alpha^m_{\rm med}$ and $\alpha_i^m$, and the training time for predicting $\alpha^m_{\rm med}$ is about 8\textperthousand~of that for predicting $\alpha_i^m$. }}
\end{rem}


\renewcommand{\algorithmicrequire}{\textbf{Input:}}
\renewcommand{\algorithmicensure}{\textbf{Output:}}
\begin{table}[htb]\small
\vspace{-0.2cm}
\caption{The Heuristic Policy}
\vspace{-1cm}
\begin{tabular}{p{16cm}}\label{T:Heuristic}
\\\hline
\end{tabular}
\vspace{-0.2cm}
\begin{algorithmic}[1]
\REQUIRE $R^m_{{i}}$, $i=1,...,N_L$, $m=1,...,M_D$, $Q_{\max}$, $\alpha^m_{\rm med}$, $m = 1,...,M_D$.
\ENSURE $P_i^m$ and $K_i^m$, $i=1,...,N_L$, $m=1,...,M_D$.
\STATE $i := 1$, $\tilde{i}:=1$, and $Q_1^m:=R_1^m$, $m=1,...,M_D$.
\WHILE{$i \leq N_L$}
\IF{$\tilde{i} < N_L +1$.}
\WHILE{$m \leq M_D$}
\IF{$\alpha^m_{i} \geq \alpha^m_{\rm med}$}
\STATE $\hat{s}^m_i := \frac{1}{\Delta T} (R^m_{\tilde{i}+1} + R^m_{\tilde{i}+2})$, $\tilde{i}:=\tilde{i}+2$, if $Q^m_i + R^m_{\tilde{i}+1} + R^m_{\tilde{i}+2} - R^m_{{i}} \leq Q_{\max}$.
\STATE $\hat{s}^m_i := \frac{1}{\Delta T} R^m_{\tilde{i}+1}$, $\tilde{i}:=\tilde{i}+1$, if $Q^m_i + R^m_{\tilde{i}+1} + R^m_{\tilde{i}+2} - R^m_{{i}} > Q_{\max}$ and $Q^m_i + R^m_{\tilde{i}+1} - R^m_{{i}} \leq Q_{\max}$.
\STATE $\hat{s}^m_i := 0 $, if $Q^m_i + R^m_{\tilde{i}+1} - R^m_{{i}} > Q_{\max}$.
\ELSE
\STATE $\hat{s}^m_i := \frac{1}{\Delta T} R^m_{i+1}$, $\tilde{i}:=\tilde{i}+1$, if $i=\tilde{i}$.
\STATE $\hat{s}^m_i := 0$, if $\tilde{i} > i$.
\ENDIF
\ENDWHILE
\STATE $\hat{s}^m_i := 0 $.
\ELSE
\STATE Solve problem \eqref{eq:Heuristic}, and obtain $P_i^m$ and $K_i^m$, $m=1,...,M_D$.
\ENDIF
\ENDWHILE
\RETURN $P_i^m$ and $K_i^m$, $i=1,...,N_L$, $m=1,...,M_D$.
\end{algorithmic}
\vspace{-0.4cm}
\begin{tabular}{p{16cm}}
\\
\hline
\end{tabular}
\vspace{-1.0cm}
\end{table}


\vspace{-2mm}\section{Simulation Results}
In this section, we evaluate the EE of the proposed ideal policy and heuristic policy.
We consider both scenarios with perfect and imperfect predictions of average channel gains.

We compare the proposed policies (with legends ``Ideal" and ``Heuristic") with four baselines.
\begin{itemize}
\item Non-predictive joint resource allocation (legend ``Baseline 1"): This is extended from a policy only considering RT users in \cite{EEECLY} by jointly optimizing average transmit power and bandwidth in each frame for VoD and RT users. For VoD users, the segments to be played in the $i$th frame are transmitted in the $i-1$th frame (i.e., $\bar{s}_{i-1}^{mn} = \frac{1}{\Delta T} R_i^m$).
\item PRA only with future average channel gains for VoD users (legend ``Baseline 2"):  This is extended from a policy  only considering VoD users in \cite{Hatem2014EE}. The unknown distances between BS and RT users are set as cell radius in all frames, and then the resource allocation for VoD and RT users are jointly optimized.
\item Decoupled PRA in two timescales (legend ``Baseline 3"):  This is extended from a two-timescale policy only considering VoD users in \cite{Hatem2014MSWiM}. The extended policy optimizes bandwidth allocation at the beginning of prediction window (equivalent to allocating the number of time slots in \cite{Hatem2014MSWiM}), where transmit power is equally allocated to all subcarriers (i.e., $P_{\max}/K_{\max}$) in order to predict average rate. In each time slot, the transmit power is allocated to subcarriers with \eqref{eq:waterfill} and \eqref{eq:waterfillRT} (similar to subcarrier allocation in \cite{Hatem2014MSWiM}).
\item Decoupled PRA for two services (legend ``Baseline 4"): This is extended from a two-timescale policy that optimizes resource allocation for file-downloading users with the residual bandwidth and power after serving RT users \cite{Yao2016Planning}. To obtain the residual resources, we first assign bandwidth to RT users with fixed transmit power on each subcarrier as $P_{\max}/K_{\max}$. In this way, the residual bandwidth is proportional to the residual power, as assumed in  \cite{Yao2016Planning}.  By solving problem \eqref{eq:aimpower} with given residual resources, the resource allocation for VoD services in two timescales is jointly optimized.
\end{itemize}

By comparing with Baseline~1, we can illustrate the gain of improving EE by harnessing future average channel gains. By comparing with Baseline~2, we can show when the future average channel gains of RT users is helpful. By comparing with Baseline 3, we can show the EE loss from decoupling the optimization in the two timescales. By comparing with Baseline 4, we can show the EE loss due to reserving resources for RT service.

Since the predicted average channel gains, denoted as $\{\hat{\alpha}_i^{m}$, $i=1,...,N_L\}$, are not error-free, the ideal policy needs to adjust resources to ensure QoS. With $\{\hat{\alpha}_i^{m}$, $i=1,...,N_L\}$, resource allocation plan $\{\hat{P}_i^{m}, \hat{K}_i^{m}, i=1,...,N_L\}$ can be obtained by solving problem \eqref{eq:aimpower}. At the beginning of the $i$th frame, if the average channel gain estimated at the BS ${\alpha}_i^{m} \ne \hat{\alpha}_i^{m}$, then we apply a simple adjustment that does not need to solve another optimization problem: the BS adjusts average transmit power according to ${{\alpha}_i^{m}}\tilde{P}^{m}_{i} = {\hat{\alpha}_i^{m}}\hat{P}_i^{m}$, and $\hat{K}_i^{m}$ does not change. Such a modified ideal policy to address prediction errors is referred to as \emph{extended ideal policy}.


\vspace{-3mm}\subsection{Simulation Setup}
For VoD services, we consider {SVC} in \cite{Patrick2012Video} (each segment includes one base layer and five enhance layers). The bit rate of each layer can be found in \cite{VideoTrace}. The average streaming rate of each VoD service is around $2$~Mbits/s. For RT services, the packets of each user arrive at the buffer of BS according to a Poisson process with average rate $\lambda_a = 500$~packets/s. The size of each packet follows an exponential distribution with average $1/{\lambda_u} = 4$~kbits/packet. Hence, the average data arrival rate of each RT service is $2$~Mbits/s.

All the users move along a road with minimal distance from BSs as $100$~m, where the distance between two adjacent BSs is $500$~m.
In the scenario with perfect prediction, the velocities of users are constant, i.e., $20$~m/s. In the scenario with imperfect prediction, the velocities are random variables, as to be detailed later. To save transmit power, each user is accessed to its nearest BS.
The path loss model is $35.3+37.6\log_{10}D$ dB, where $D$ is the distance in meters between a user and its accessed BS in a frame. The circuit powers of different components are obtained from those  measured in the year of $2012$ in \cite{Claude2012Flexible}, where the scaling law in \cite{Claude2014Modeling} is further applied to predict $P_c$ and $P_0$ in $2020$. The prediction window is with duration $N_L \Delta T = 60$~s. The results for the scenarios with perfect and imperfect prediction are respectively obtained from 100 and 1000 simulation trails. In each trail, the user trajectory in the prediction window, the packet arrival and packet size of RT services, and Rayleigh fading channels are randomly generated. Matlab is used for the simulation. The simulation parameters are listed in Table \ref{T:parameter}. This setup will be used in the sequel unless otherwise specified.
\begin{table}[htbp]
\small
\renewcommand{\arraystretch}{1.3}
\caption{List of Simulation Parameters \cite{Claude2012Flexible,Claude2014Modeling}}\vspace{-0.6cm}
\begin{center}\vspace{-0.4cm}\label{T:parameter}
\begin{tabular}{|p{5.4cm}|p{2.0cm}||p{5.4cm}|p{1.9cm}|}
  \hline
  Maximal transmit power $P_{\max}$ & $40.0$ W &
  Total number of subcarriers $K_{\max}$ & $512$\\\hline
  Bandwidth of each subcarrier $B$ &$15$ kHz&
  Power amplifier efficiency $\rho$ & $38.8$~\% \\\hline
  Single-sided noise spectral density $N_0$ & -173~dBm/Hz&
  Fixed circuit power consumption $P_0$ & $136$~mW/MHz \\\hline
  Circuit power consumption for one subcarrier $P_c$ & $72$~mW/MHz &
  Duration of each frame $\Delta T$ and each time slot & $1$~s and $5$~ms\\\hline
\end{tabular}
\end{center}
\vspace{-0.8cm}
\end{table}

\vspace{-2mm}\subsection{Perfect Prediction of Average Channel Gains}
\begin{figure}[htbp]  
\vspace{-0.4cm}
\centering
\subfigure[{EE v.s. ratios of the number of VoD users.}]{
\label{fig:EEMD} 
\includegraphics[width=0.43\textwidth]{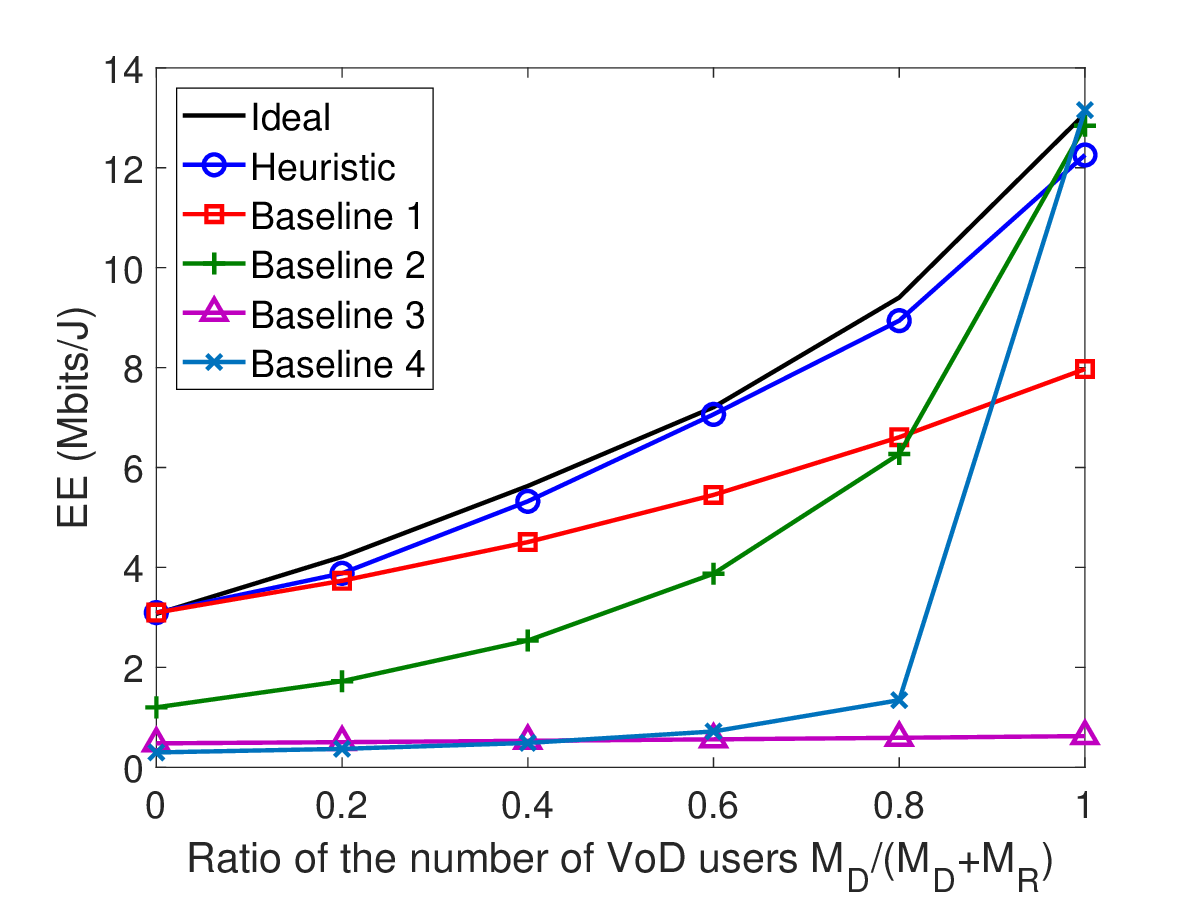}}
\subfigure[{EE v.s. streaming rate of VoD user, $M_D=M_R=1$.}]{
\label{fig:EERatioD} 
\includegraphics[width=0.43\textwidth]{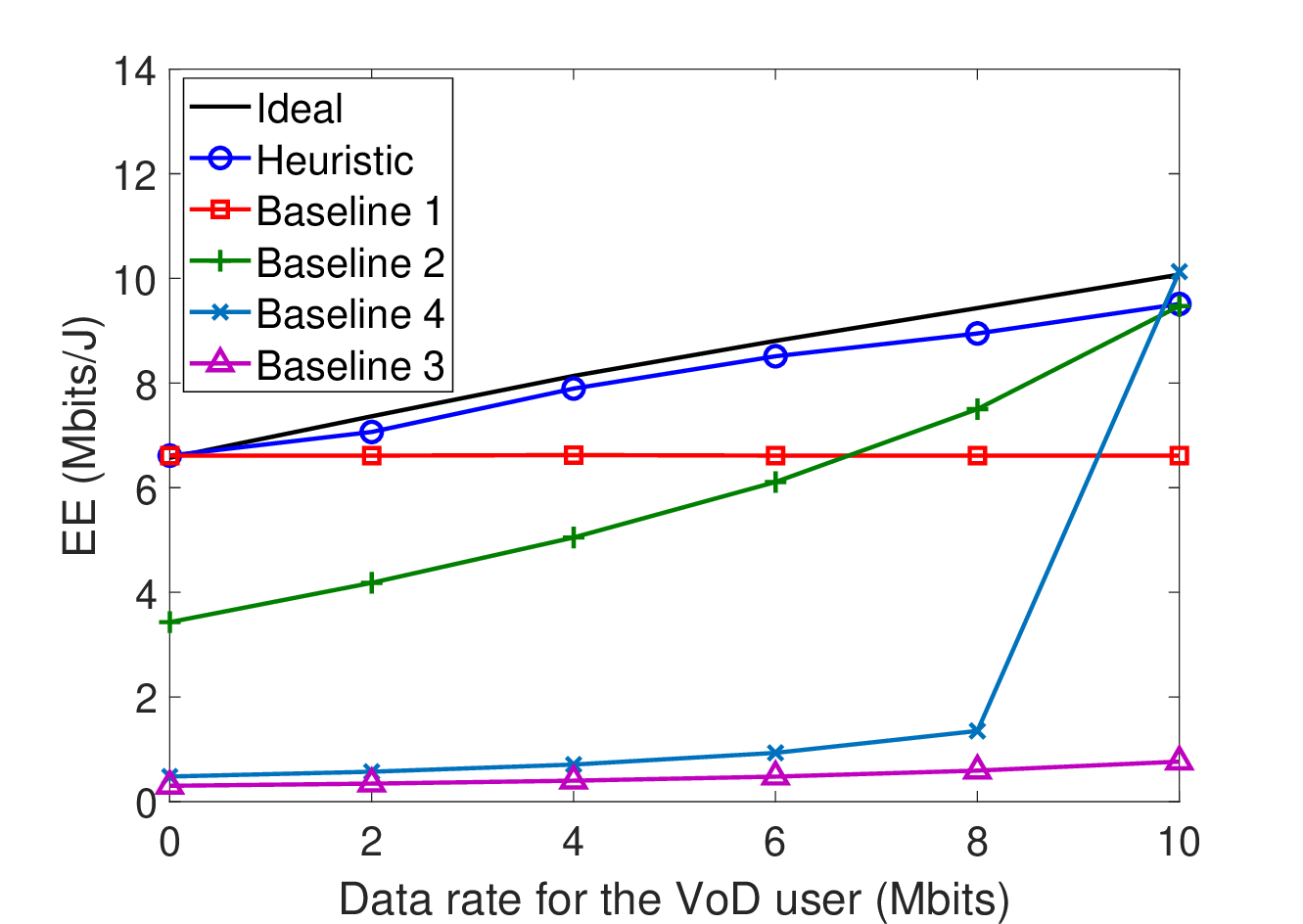}}\vspace{-0.2cm}
\caption{EE achieved by different policies.}
 \label{fig:EER} 
\vspace{-0.4cm}
\end{figure}

The EE achieved by different policies is shown in Fig. \ref{fig:EER}. In Fig. \ref{fig:EEMD}, the total number of users is fixed as $M_D+M_R = 5$. In Fig. \ref{fig:EERatioD}, the sum data rate required by all users are fixed as ${\mathbb{E}}(R^m_i/\Delta T) + \lambda_a/\lambda_u = 10$~Mbits, where the arrival data rate of RT user (or the streaming rate of VoD user) varies. We can see that when there is no VoD user, the achieved EE of the ideal policy and Baseline 1 are identical. The results are consistent with Observation 2, i.e., predicting average channel gains cannot help improve the EE of the system only with RT users. When there are both VoD and RT users, the achieved EE of the ideal policy could be $50 \sim 100$\% higher than the EE achieved by the baselines. The achieved EE of Baseline 2 is lower than Baseline 1 when the number (or arrival data rate) of RT users is large because the resources reserved for the RT users are too conservative. The achieved EEs of Baseline 3 and Baseline 4 are much lower than Baseline 1, which is a non-predictive policy that jointly optimizes resource allocation for the two types of services. This result suggests the necessity of the joint optimization for two types of services in the two timescales. Since Baselines 3 and 4 perform the worst almost in all cases, we no longer provide their performance in the sequel unless necessary. The heuristic policy performs closely to the ideal policy.

\begin{table}[htbp]
\small
\renewcommand{\arraystretch}{1.3}
\caption{Numbers of Layers of SCV with Different Number of Users}\vspace{-0.6cm}
\begin{center}\vspace{-0.2cm}\label{T:quality}
\begin{tabular}{|p{2.5cm}|p{1cm}|p{1cm}|p{1cm}|p{1cm}|p{1cm}|}
  \hline
  $M_D=M_R$ & $\leq 8$ & $9$ & $10$ & $11$ & $\geq 12$ \\\hline
  Ideal & $5$ & $4.9667$ & $4.8833$ & $4.6500$ & NA \\\hline
  Heuristic & $5$ & $4.9167$ & $4.7833$ & $4.6167$ & NA \\\hline
  Baseline 1& $5$ & $4.8333$ & $4.7500$ & $4.5800$ & NA \\\hline
  Baseline 2 & $5$ & $4.4833$ & $4.0800$ & NA & NA \\\hline
\end{tabular}
\end{center}
\vspace{-1.2cm}
\end{table}

To show the throughput of the considered network in terms of maximal total number of RT and VoD users without stalling, we provide the video quality with different numbers of users in Table \ref{T:quality}, which provides the average number of enhanced layers transmitted to VoD users. To obtain the results, we first set the video quality at level 5 (all the 5 enhanced layers are transmitted) and find the solutions with different policies. If the problem is infeasible, we reduce the video quality of VoD users to a lower level until the problem is feasible. Finally, we calculate the average video quality of VoD users in $100$ minutes. ``NA" means that at least in one frame, the QoS of RT users cannot be satisfied or the data in base layer cannot be transmitted to VoD users (i.e., playback interruption occurs). Due to the lack of space, we do not show the performance when stalling occurs, which is acceptable in practice. We can see that the maximal total number of users that the system can support with ensured QoS is 20 (i.e., $M_D=M_R=10$) if using Baseline 2, and is 22 if using other three policies. Again, the heuristic policy  performs closely to the ideal policy in term of the video quality. Since the EE can be improved only when {the} traffic load is not high (say $M_D, M_R<10$ for the considered setup), in the sequel we only consider the scenario where the QoS of all VoD users can be ensured.

\begin{figure}[!htb]
	\begin{minipage}[t]{0.48\linewidth}
		\centering
		\vspace{-3mm}
		\includegraphics[width= 0.99\textwidth]{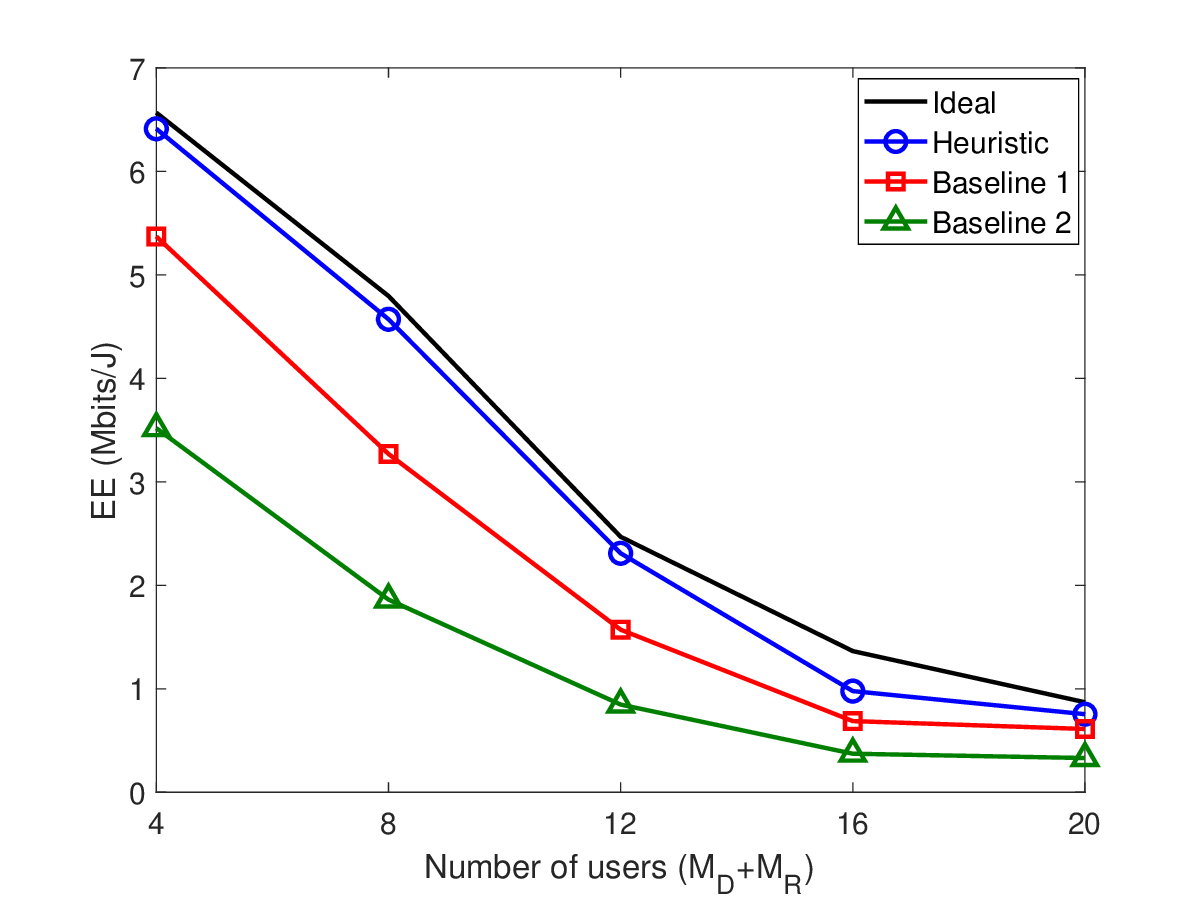}
		\vspace{-5mm}
		\caption{EE v.s. total number of users, where $M_D = M_R$.}
		\label{fig:EEuser}
	\end{minipage}%
	\begin{minipage}[t]{0.48\linewidth}
		\centering
		\vspace{-3mm}
		\includegraphics[width= 0.99\textwidth]{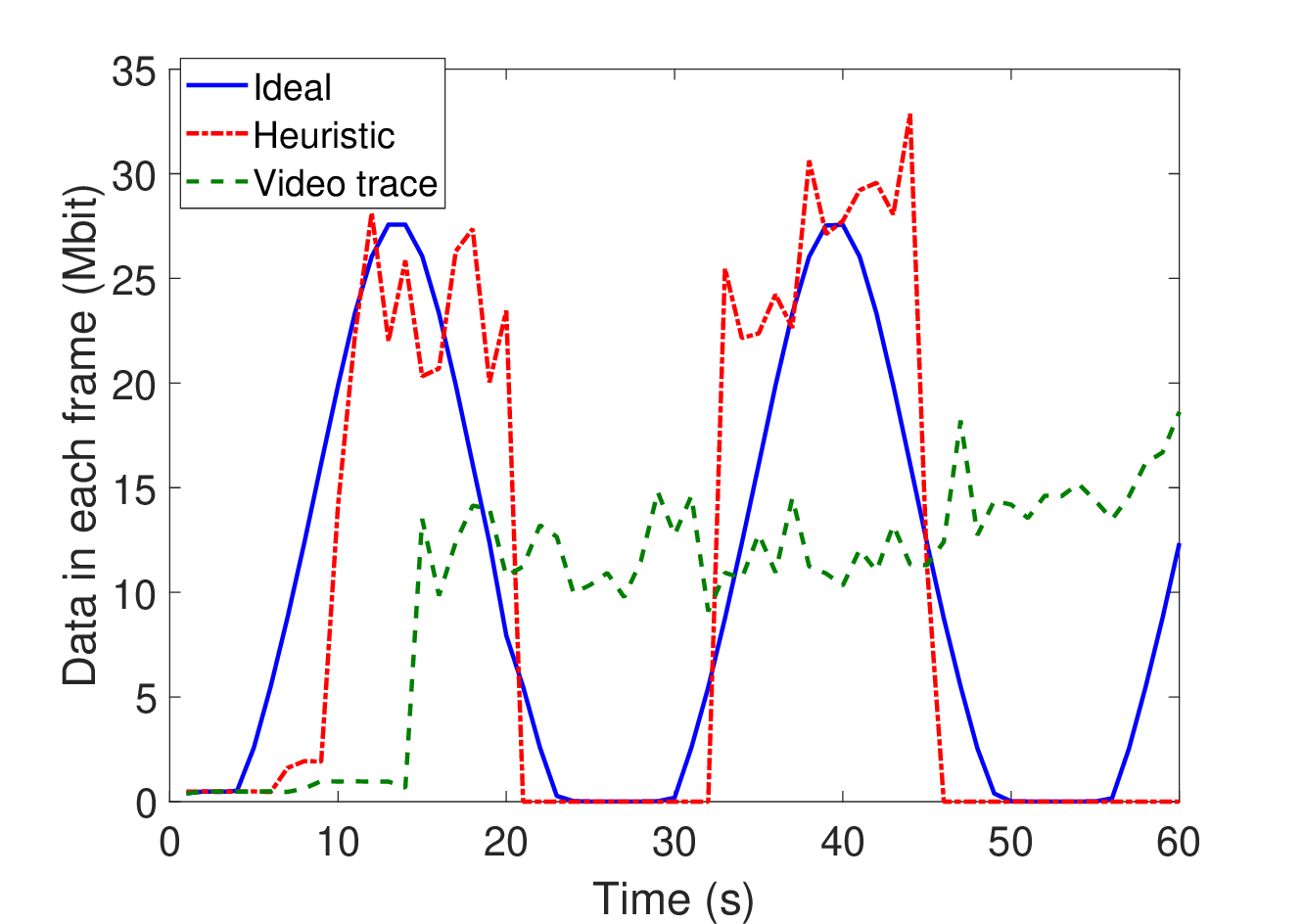}
		\vspace{-7mm}
		\caption{Data transmitted in different frames, where the average streaming rate of the VoD service and the average data arrival rate of the RT service are set as $10$~Mbps.}
		\label{fig:rate}
	\end{minipage}
	\vspace{-8mm}
\end{figure}



The EE of the system supporting different traffic loads is shown in Fig. \ref{fig:EEuser}. The results show that EE achieved by the ideal and heuristic policies are much higher than that achieved by the baselines when the number of users is small. When the number of users approaches to the maximal number of users that the system can support, the EE achieved by different policies are almost identical. This is because when the network is fully loaded, the BSs need to serve the users with maximal transmit power and bandwidth, and hence there is no chance to save energy. 

To understand the behavior of the heuristic policy, we consider a simplified scenario with one VoD user and one RT user. The amount of data transmitted in different frames and the amount of video data played in each frame (with legend ``video trace") are shown in Fig. \ref{fig:rate}. The results show that at the first several frames in the prediction window (i.e., the beginning of the service), the data amount transmitted in each frame equals to the data amount to be played in the next frame. After the first several seconds, both ideal and heuristic policies transmit data when the large-scale channel gains are good. There is no stalling since the video segments are transmitted before playback.

\begin{figure}[htbp]  
\vspace{-0.2cm}
\centering
\subfigure[{Bandwidth allocation.}]{
\label{fig:KI} 
\includegraphics[width=0.42\textwidth]{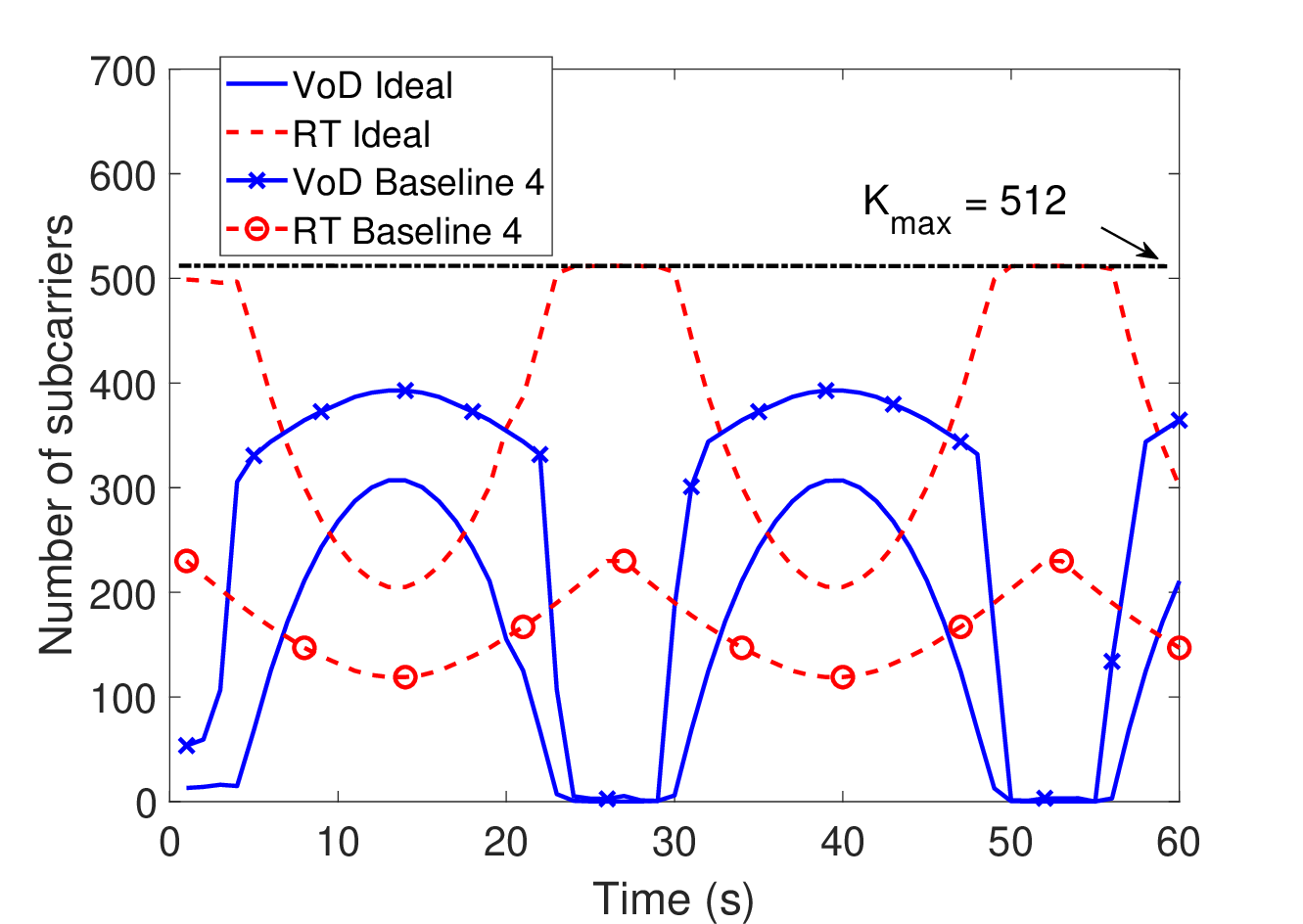}}
\subfigure[{Average power allocation.}]{
\label{fig:PtI} 
\includegraphics[width=0.42\textwidth]{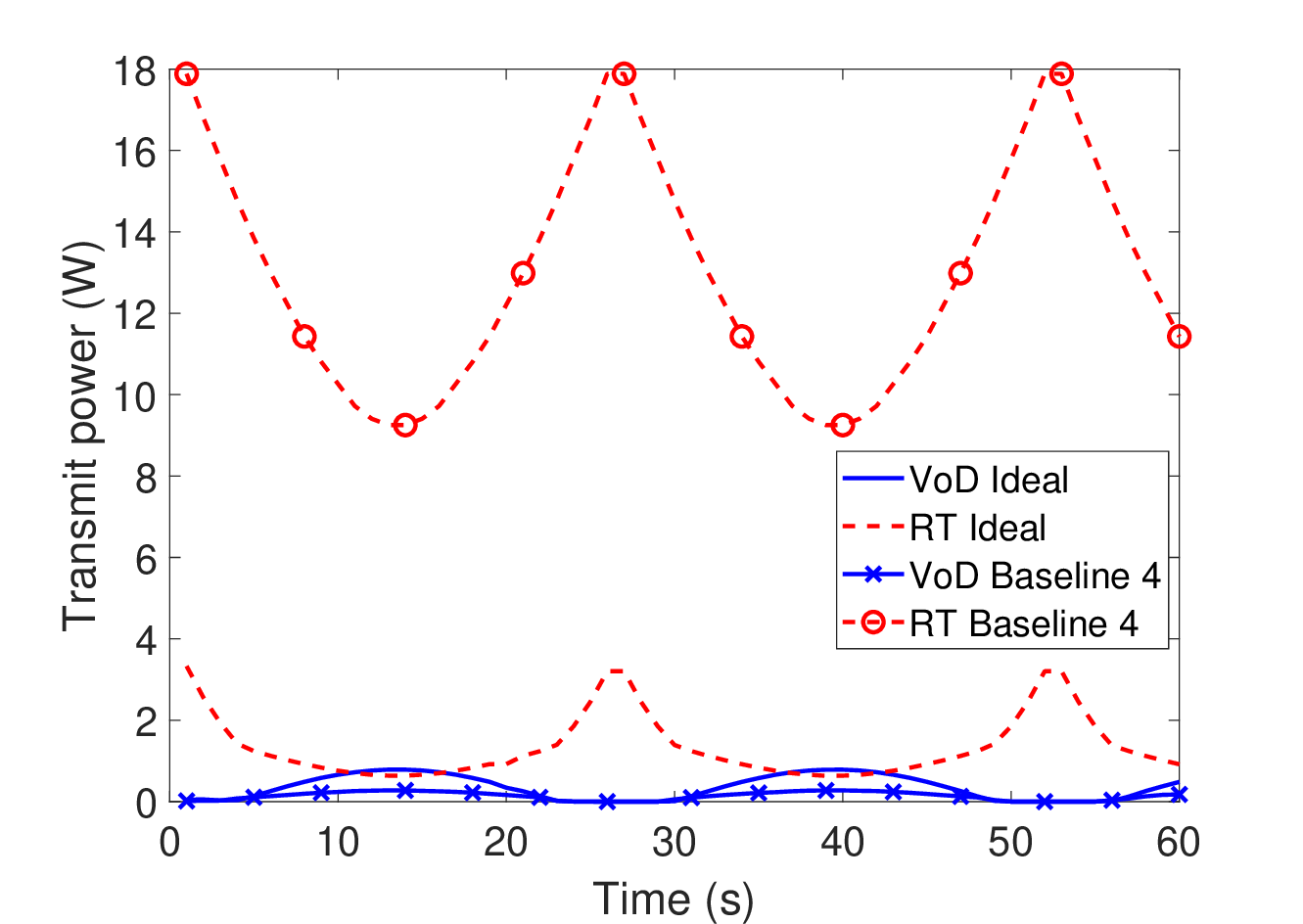}}\vspace{-0.2cm}
\caption{Coupled resource allocation between VoD and RT services, where the average streaming rate of the VoD service and the average data arrival rate of the RT service are set as $10$~Mbps. We use solid and dash lines and further add cross and dot marks in these lines to distinguish different results.}
 \label{fig:RA} 
\vspace{-0.8cm}
\end{figure}

To understand why the resource allocation between VoD and RT services should be jointly optimized, we again consider the simple scenario with one VoD user and one RT user. The average transmit power and bandwidth assigned by the ideal policy and Baseline 4 in each frame to each service are shown in Fig. \ref{fig:RA}. The results show that if the resource allocation is jointly optimized, more subcarriers will be allocated to the RT users if the average data rate of a VoD user (say the $m$th user), $\bar{s}_i^m$, is zero in a frame (e.g., the user is located in cell-edge). However, with Baseline 4, the BS first assigns subcarriers (i.e., reserves resources) to RT service without considering the bandwidth required by VoD service. To leave some bandwidth for VoD service, the BS will not assign all the subcarriers to the RT service. With less bandwidth, more transmit power is consumed by the RT service, and hence the EE of the system is low.

\vspace{-5mm}
\subsection{Imperfect Prediction of Large-scale Channel Gains}
The prediction errors may come from many sources such as erroneous mobility route prediction, inaccurate velocity prediction, user location estimation and constructed radio map. Here we take the velocity prediction error as an example to illustrate the impact of imperfect prediction, since it leads to the errors on average channel gains accumulative with frames and hence causes more severe performance degradation than other types of prediction errors.

Markov chain is widely used to model the mobility of vehicles \cite{bui2016anticipatory}. We use a discrete time Markov chain to characterize the velocity of each user. Specifically, the velocity of each user lies in $\mathcal{V}= \{v_{1},v_2,...,v_U\}$, where $\Delta v \triangleq v_{u+1}-v_u = 1$~m/s,  $v_1 = 0$~m/s, and $v_U = 30$~m/s, and the velocities are constant within each frame of duration $\Delta T$. With this model, the velocity of a user may vary from $0\sim 30$~m/s in the prediction window. Denote the velocity of the $m$th user in the $i$th frame as $V^m_i$. We set $\Delta v/\Delta T$ equal to the maximal acceleration of vehicles (e.g., $1$~$\text{m}/\text{s}^2$ \cite{TETC2013}). The velocity can only transit between adjacent states (i.e., it can change $\Delta v$ after $\Delta T$). The $U\times U$ transition matrix of the Markov chain is denoted as ${\bf{U}}$, where $u_{i,j}$ is the probability that the velocity transits from $v_{{i}}$ to $v_{{j}}$. For the considered scenario, $u_{11}=u_{U,U} = 1-q$, $u_{i,i} = 1-2q$, $i \ne 1, U$, and $u_{i,i+1} = u_{i+1,i} = q, i=1,...,U-1$. When $q=0$, the velocity is constant and always equals to the initial value. By increasing the value of $q$, the prediction errors of velocities in the upcoming $N_L$ frames increase. We do not study how to predict the trajectory of each user, and apply a simple way to illustrate the performance of different policies. Specifically, the predicted locations of the user are obtained by assuming that each user travels along the predicted route with the initial velocity, setting as $20$~m/s.\footnote{According to simulations, the uncertainty of velocity modeled in the sequel will lead to $200\sim300$ \% prediction errors on average channel gains at the end of a prediction window with $60$ seconds duration.} The initial positions of the users are uniformly distributed in the first cell. The median of each VoD user is computed from the predicted average
channel gains of the user in the window.

Since EE can be improved evidently when the traffic load is light, we set $M_D = M_R = 5$, which is half of the maximal number of users that can be supported by the BSs.

\vspace{-0.0cm}
\begin{figure}[htbp]
        \centering
        \begin{minipage}[t]{0.42\textwidth}
        \includegraphics[width=1\textwidth]{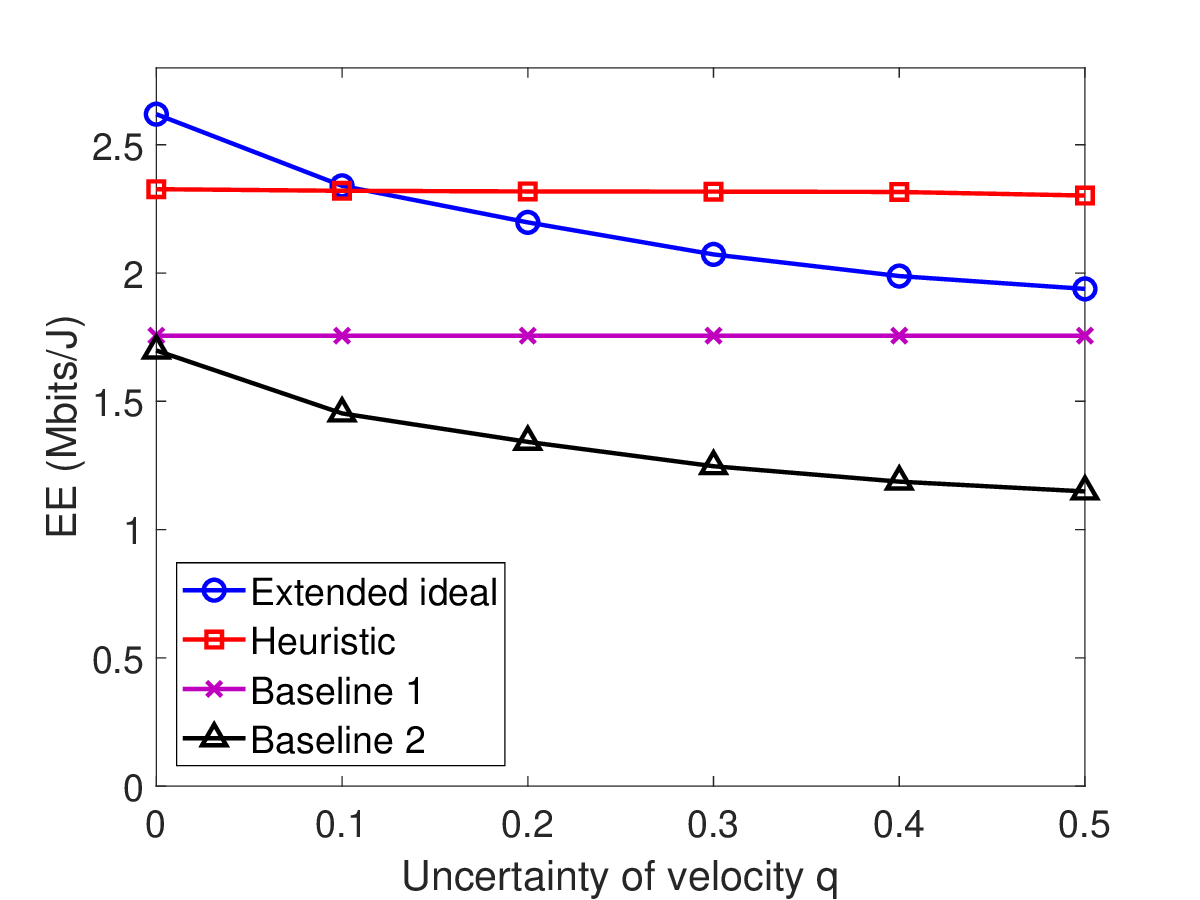}
        \end{minipage}\vspace{-5mm}
        \caption{EE v.s. uncertainty of velocity.}
        \label{fig:EEq}\vspace{-0.5cm}
\end{figure}

The EE achieved by different policies\footnote{We do not compare with the robust policies in \cite{Ramy2017TWC,Ramy2018GCT}. This is because existing methods can only reformulate non-deterministic linear constraints as deterministic constraints \cite{Aharon2009Robust}, but the constraint in \eqref{eq:Cst} is non-linear. } is shown in Fig. \ref{fig:EEq}, where the qualities of all the VoD users are the same (all six layers are transmitted before playback). Since the ideal policy is optimized under the assumption of perfect prediction of average channel gains, it is no longer optimal when the prediction is non-perfect. When $q$ is large, the heuristic policy outperforms the ideal policy. Even when $q=0.5$, which leads to over 300\% prediction errors on the average channel gains at the end of the prediction window, the EE loss of heuristic policy is negligible.

\vspace{-3mm}\section{{Conclusion}}
In this paper, we demonstrated the gain in maximizing EE of an OFDMA system with both VoD and RT services by harnessing the prediction of average channel gains in a time window, and investigated which kinds of future information
are needed to improve EE. To this end, an optimal two-timescale PRA policy was obtained, which needs fine-grained prediction as in existing works. By analyzing the optimal policy, we found that using instantaneous channel gains in future time slots can not improve EE. To reduce the training cost incurred by the fine-grained prediction, a heuristic policy was proposed. This policy only needs the median of future average channel gains of VoD users, and jointly optimizes the average power and bandwidth for RT and VoD users in each frame. Simulation results showed that EE can be improved remarkably by jointly optimizing resource allocation for the two types of services in the two timescales, which suggests the necessity of sharing resources among different services. The heuristic policy performs closely to the optimal policy if the prediction is error-free, and outperforms the optimal policy if the prediction is with large uncertainty. This demonstrates that the EE gain over non-predictive policies can be achieved by non-perfect coarse-grained prediction.



\appendices
\section{Proof of Proposition \ref{P:optimal}}
\label{App:optimal}
\renewcommand{\theequation}{A.\arabic{equation}}
\setcounter{equation}{0}
\begin{proof}
To prove Proposition \ref{P:optimal}, we first prove that $f_D^w\left(\frac{\bar{P}^m_i}{K^m_i},g\right)$ is the optimal power allocation policy for VoD services. For arbitrary power allocation policy for RT services $f'_R\left({\bar{P}^m_i},{K^m_i},g\right)$, the optimal solutions of problem \eqref{eq:aimenergy} with policies $f_D^w\left(\frac{\bar{P}^m_i}{K^m_i},g\right)$ and $f'_D\left({\bar{P}^m_i},{K^m_i},g\right)$ are denoted as $\{\tilde{{P}}^m_i, \tilde{K}^m_i, m = 1,...,M_D+M_R, i = 1,...,N_L\}$ and $\{{{P}^m_i}', {{K}^m_i}', m = 1,...,M_D+M_R, i = 1,...,N_L\}$, respectively. Then, we need to prove $E^*_{\rm ave}\left(f_D^w, f'_R\right) \leq E^*_{\rm ave}\left(f'_D, f'_R\right)$, where
\begin{align}
E^*_{\rm ave}\left(f_D^w, f'_R\right) = \sum\limits_{m = 1}^{{M_D+M_R}}{\sum\limits_{i = 1}^{{N_L }} {\left( \frac{1}{\rho }{\tilde{P}_i^{m}} + {P_c}{\tilde{K}_i^{m}}\right)}}; E^*_{\rm ave}\left(f'_D, f'_R\right) = \sum\limits_{m = 1}^{{M_D+M_R}}{\sum\limits_{i = 1}^{{N_L}} {\left( {\frac{1}{\rho }{{{P}^m_i}'} + {P_c}{{{K}^m_i}'}} \right)}}.\nonumber
\end{align}
Denote the average service rates achieved by the power allocation policies $f'_D\left({\bar{P}^m_i},{K^m_i},g\right)$ with resource allocation planning $\{{{P}^m_i}', {{K}^m_i}'\}$ as ${{s}^m_i}'$, $m = 1,...,M_D+M_R$, $i=1,...,N_L$.

To prove $E^*_{\rm ave}\left(f_D^w, f'_R\right) \leq E^*_{\rm ave}\left(f'_D, f'_R\right)$, we need the following result: the water-filling policy $f_D^w\left(\frac{\bar{P}^m_i}{K^m_i},g\right)$ can minimize $\bar{P}^m_i$ with given $K^m_i$ and average service rate $\bar{s}^m_i$ \cite{WirelessCom}.
According to this result, given ${{K}^m_i}'$ and ${{s}^m_i}'$, $i=1,...,N_L$, the average transmit power is minimized with $f_D^w\left(\frac{\bar{P}^m_i}{K^m_i},g\right)$. Denote the related minimal average transmit power for the $m$th user in the $i$th frame as $\min(\bar{P}^{m}_i)$. Then, $\min(\bar{P}^{m}_i) \leq {{P}^m_i}'$, $m = 1,...,M_D, i=1,...,N_L$. Hence

\begin{align}
\sum\limits_{m = 1}^{{M_D }}{\sum\limits_{i = 1}^{{N_L }} {\left[ \frac{1}{\rho }{\min({\bar P}^{m}_i)} + {P_c}{{{K}^m_i}'}\right]}}+ \sum\limits_{m = M_D}^{{M_D+M_R}}{\sum\limits_{i = 1}^{{N_L }} {\left( {\frac{1}{\rho }{{{P}^m_i}'} + {P_c}{{{K}^m_i}'}}\right)}}
\leq \sum\limits_{m = 1}^{{M_D+M_R}}{\sum\limits_{i = 1}^{{N_L}} {\left( {\frac{1}{\rho }{{{P}^m_i}'} + {P_c}{{{P}^m_i}'}} \right)}}. \label{eq:A1}
\end{align}

Moreover, with $f_D^w\left(\frac{\bar{P}^m_i}{K^m_i},g\right)$, the optimal resource allocation plan is $\{\tilde{P}_i^{m}, \tilde{K}_i^{m}$, $m = 1$,...,$M_D+M_R$, $i = 1$,...,$N_L\}$. Thus, $\sum\limits_{m = 1}^{{M_D+M_R}}{\sum\limits_{i = 1}^{{N_L }} {\left( \frac{1}{\rho }{\tilde{P}_i^{m}} + {P_c}{\tilde{K}_i^{m}}\right)}} \leq \sum\limits_{m = 1}^{{M_D }}{\sum\limits_{i = 1}^{{N_L }} {\left[ \frac{1}{\rho }{\min({\bar P}^{m}_i)} + {P_c}{{{K}^m_i}'}\right]}}+ \sum\limits_{m = M_D}^{{M_D+M_R}}$ ${\sum\limits_{i = 1}^{{N_L }} {( {\frac{1}{\rho }{{{P}^m_i}'} + {P_c}{{{K}^m_i}'}})}}$.
Further considering \eqref{eq:A1}, we have $E^*_{\rm ave}\left(f_D^w, f'_R\right) \leq E^*_{\rm ave}\left(f'_D, f'_R\right)$.

Similarly, given power allocation policy for VoD services $f_D^w\left(\frac{\bar{P}^m_i}{K^m_i},g\right)$, we can prove that
$E^*_{\rm ave}\left(f_D^w, f_R^w\right) \leq E^*_{\rm ave}\left(f_D^w, f'_R\right).\label{eq:optimalRT}$
The proof is omitted for conciseness. Therefore, we can obtain that
$E^*_{\rm ave}\left(f_D^w, f_R^w\right) \leq E^*_{\rm ave}\left(f_D^w, f'_R\right) \leq E^*_{\rm ave}\left(f'_D, f'_R\right)$.
\end{proof}

\section{Proof of Property \ref{P:EC}}
\label{App:PEC}
\renewcommand{\theequation}{B.\arabic{equation}}
\setcounter{equation}{0}
\begin{proof}
The proof of the convexity of \eqref{eq:CVOD} is shown in the conference version \cite{ICCC2015}.

We only prove the convexity of \eqref{eq:CRT}. The left-hand side of \eqref{eq:CRT} is the perspective of $- \frac{1}{{{\theta ^m}\tau }}\ln \left[ {F_R}\left( \bar{P}_{S_i}^m \right)\right] $, where $\bar{P}_{S_i}^m = {\frac{{\bar P_i^m}}{{K_i^m}}}$. To prove that the left-hand side of \eqref{eq:CRT} is jointly concave in ${{\bar P_i^m}}$ and ${{K_i^m}}$, we only need to prove that $- \frac{1}{{{\theta ^m}\tau }}\ln \left[ {F_R}\left( \bar{P}_{S_i}^m \right)\right] $ is concave in $\bar{P}_{S_i}^m$. For notational simplicity, we omit indices $m$ and $i$ of all the variables in this appendix. In the sequel, we will prove that
\begin{align}
\frac{{{d^2}\left[ {\ln {F_R}\left( {{{\bar P}_S}} \right)} \right]}}{{{d^2}{{\bar P}_S}}} = \frac{{{F_R}\left( {{{\bar P}_S}} \right)\frac{{{d^2}{F_R}\left( {{{\bar P}_S}} \right)}}{{d\bar P_S^2}} - {{\left[ {\frac{{d{F_R}\left( {{{\bar P}_S}} \right)}}{{d{{\bar P}_S}}}} \right]}^2}}}{{{{\left[ {{F_R}\left( {{{\bar P}_S}} \right)} \right]}^2}}} > 0. \label{eq:lnFP}
\end{align}

Substituting \eqref{eq:waterfillRT} into \eqref{eq:FRT}, we can obtain that
\begin{align}
{F_R}\left( {{{\bar P}_S}} \right) = 1 - {e^{ - \nu }} + {\nu ^{\frac{\beta }{{\beta  + 1}}}}\int_\nu ^\infty  {{g^{ - \frac{\beta }{{\beta  + 1}}}}{e^{ - g}}{\rm d}g} \label{eq:FP}
\end{align}
where the relation between ${\nu}$ and ${\bar P}_S$ can be obtained from \eqref{eq:cutoffRT}. Then, ${F_R}\left( {{{\bar P}_S}} \right)$ can be regarded as a composition function ${F_R}\left[{\nu}\left( {{{\bar P}_S}} \right)\right]$, and thus
\begin{align}
\frac{{d{F_R}\left[ {\nu \left( {{{\bar P}_S}} \right)} \right]}}{{d{{\bar P}_S}}} = \frac{{d{F_R}}}{{d\nu }}\frac{{d\nu }}{{d{{\bar P}_S}}},\quad
\frac{{{d^2}{F_R}\left[ {\nu \left( {{{\bar P}_S}} \right)} \right]}}{{d\bar P_S^2}} = \frac{{{d^2}{F_R}}}{{d{{\nu}^2}}}{\left( {\frac{{d\nu }}{{d{{\bar P}_S}}}} \right)^2} + \frac{{d{F_R}}}{{d\nu }}\frac{{{d^2}\nu }}{{d\bar P_S^2}}\label{eq:FPd2}.
\end{align}
From \eqref{eq:cutoffRT}, we can derive the relation between ${\bar P}_S$ and ${\nu}$, i.e., $\frac{{d{{\bar P}_S}}}{{d\nu }} =  - \frac{{\phi \sigma _0^2}}{{\alpha \left( {\beta  + 1} \right)}}{\nu ^{ - \frac{{\beta  + 2}}{{\beta  + 1}}}}\int_\nu ^\infty  {{g^{ - \frac{\beta }{{\beta  + 1}}}}{e^{ - g}}{\rm d}g}$. According to the characteristic of inverse function ( i.e., $\frac{{d{{\bar P}_S}}}{{d{\nu}}} \frac{{d{\nu}}}{{d{{\bar P}_S}}} =1$ at any point $({\nu},\bar {P}_S)$ ), we can derive $\frac{{d\nu }}{{d{{\bar P}_S}}}$ from \eqref{eq:cutoffRT}, i.e.,
\begin{align}
\frac{{d\nu }}{{d{{\bar P}_S}}} =  - \frac{{\alpha \left( {\beta  + 1} \right)}}{{\phi \sigma _0^2}}{\nu ^{\frac{{\beta  + 2}}{{\beta  + 1}}}}\frac{1}{{\int_\nu ^\infty  {{g^{ - \frac{\beta }{{\beta  + 1}}}}{e^{ - g}}{\rm d}g} }}.\label{eq:vpd1}
\end{align}
From $\frac{{d^2{\nu}}}{{d{{\bar P}_S}}^2} = \frac{d\frac{{d{\nu}}}{{d{{\bar P}_S}}}}{d{\nu}}\frac{d{\nu}}{d\bar{P}_S}$, we can derive that
\begin{align}
\frac{{{d^2}\nu }}{{d\bar P_S^2}} = {\left( {\frac{\alpha }{{\phi \sigma _0^2}}} \right)^2}\left[ {\left( {\beta  + 2} \right){\nu ^{\frac{1}{{\beta  + 1}}}}\frac{1}{\varphi } + \left( {\beta  + 1} \right){\nu ^{\frac{2}{{\beta  + 1}}}}{e^{ - \nu }}\frac{1}{{{\varphi ^2}}}} \right]\left[ {\left( {\beta  + 1} \right){\nu ^{\frac{{\beta  + 2}}{{\beta  + 1}}}}\frac{1}{\varphi }} \right],\label{eq:vpd2}
\end{align}
where $\varphi = {\int_\nu ^\infty  {{g^{ - \frac{\beta }{{\beta  + 1}}}}{e^{ - g}}{\rm d}g} }$. From \eqref{eq:FP}, we have
\begin{align}
\frac{{d{F_R}}}{{d\nu }} = \frac{\beta }{{\beta  + 1}}{\nu ^{ - \frac{1}{{\beta  + 1}}}}\varphi,\quad \frac{{{d^2}{F_R}}}{{d{\nu ^2}}} =  - \frac{\beta }{{{{\left( {\beta  + 1} \right)}^2}}}{\nu ^{ - \frac{{\beta  + 2}}{{\beta  + 1}}}}\varphi  - \frac{\beta }{{\beta  + 1}}{\nu ^{ - 1}}{e^{ - \nu }}\label{eq:Fvd2}.
\end{align}
Substituting \eqref{eq:vpd1}, \eqref{eq:vpd2} and \eqref{eq:Fvd2} into \eqref{eq:FPd2}, we can derive that
\begin{align}
\frac{{d{F_R}\left[ {\nu \left( {{{\bar P}_S}} \right)} \right]}}{{d{{\bar P}_S}}} =  - \frac{\alpha }{{\phi \sigma _0^2}}\beta \nu, \quad
\frac{{{d^2}{F_R}\left[ {\nu \left( {{{\bar P}_S}} \right)} \right]}}{{d\bar P_S^2}} = {\left( {\frac{\alpha }{{\phi \sigma _0^2}}} \right)^2}\beta \left( {\beta  + 1} \right){\nu ^{\frac{{\beta  + 2}}{{\beta  + 1}}}}\frac{1}{\varphi }\label{eq:FvPd2}.
\end{align}
Upon substituting \eqref{eq:FvPd2}, the numerator of \eqref{eq:lnFP} can be derived as follows,
\begin{align}
\left( {1 - {e^{ - \nu }}} \right){\left( {\frac{\alpha }{{\phi \sigma _0^2}}} \right)^2}\beta \left( {\beta  + 1} \right){\nu ^{\frac{{\beta  + 2}}{{\beta  + 1}}}}\frac{1}{\varphi } + {\left( {\frac{\alpha }{{\phi \sigma _0^2}}} \right)^2}\beta {\nu ^2}.\label{eq:numerator}
\end{align}
Since $\varphi = {\int_\nu ^\infty  {{g^{ - \frac{\beta }{{\beta  + 1}}}}{e^{ - g}}{\rm d}g} } > 0$, $\beta = \frac{\theta \tau B}{\ln 2} >0$, \eqref{eq:numerator} is positive, and hence we have \eqref{eq:lnFP}.
\end{proof}

\bibliographystyle{IEEEtran}
\bibliography{ref}

\begin{thebibliography}{10}
\providecommand{\url}[1]{#1}
\csname url@samestyle\endcsname
\providecommand{\newblock}{\relax}
\providecommand{\bibinfo}[2]{#2}
\providecommand{\BIBentrySTDinterwordspacing}{\spaceskip=0pt\relax}
\providecommand{\BIBentryALTinterwordstretchfactor}{4}
\providecommand{\BIBentryALTinterwordspacing}{\spaceskip=\fontdimen2\font plus
\BIBentryALTinterwordstretchfactor\fontdimen3\font minus
  \fontdimen4\font\relax}
\providecommand{\BIBforeignlanguage}[2]{{%
\expandafter\ifx\csname l@#1\endcsname\relax
\typeout{** WARNING: IEEEtran.bst: No hyphenation pattern has been}%
\typeout{** loaded for the language `#1'. Using the pattern for}%
\typeout{** the default language instead.}%
\else
\language=\csname l@#1\endcsname
\fi
#2}}
\providecommand{\BIBdecl}{\relax}
\BIBdecl

\bibitem{ICCC2015}
C.~She and C.~Yang., ``Context aware energy efficient optimization for video
  on-demand service over wireless networks,'' in \emph{Proc. IEEE ICCC}, 2015.

\bibitem{NB2018}
N.~Bui and J.~Widmer, ``Data-driven evaluation of anticipatory networking in
  {LTE} networks,'' \emph{IEEE Trans. on Mobile Comput.}, vol.~17, no.~10, pp.
  2252--2265, Oct. 2018.

\bibitem{Shunqing2017Fundamental}
S.~Zhang, Q.~Wu, S.~Xu, and G.~Y. Li, ``Fundamental green tradeoffs:
  Progresses, challenges, and impacts on 5{G} networks,'' \emph{IEEE Commun.
  Surveys Tuts.}, vol.~19, no.~1, pp. 33--56, 2017.

\bibitem{EEECLY}
C.~Xiong, G.~Y. Li, Y.~Liu, Y.~Chen, and S.~Xu, ``Energy-efficient design for
  downlink {OFDMA} with delay-sensitive traffic,'' \emph{IEEE Trans. Wireless
  Commun.}, vol.~12, no.~6, pp. 3085--3095, Jun. 2013.

\bibitem{Amir2013Energy}
A.~Helmy, L.~Musavian, and T.~Le-Ngoc, ``Energy-efficient power adaptation over
  a frequency-selective fading channel with delay and power constraints,''
  \emph{IEEE Trans. Wireless Commun.}, vol.~12, no.~9, pp. 4529 -- 4541, Sep.
  2013.

\bibitem{Changyang2015Tcom}
C.~She, C.~Yang, and L.~Liu, ``Energy-efficient resource allocation for
  {MIMO}-{OFDM} systems serving random sources with statistical {Q}o{S}
  requirement,'' \emph{IEEE Trans. Commun.}, vol.~63, no.~11, pp. 4125--4141,
  Nov. 2015.

\bibitem{yang2018energy}
Z.~Yang, W.~Xu, Y.~Pan, C.~Pan, and M.~Chen, ``Energy efficient resource
  allocation in machine-to-machine communications with multiple access and
  energy harvesting for {I}o{T},'' \emph{IEEE Internet of Things J.}, vol.~5,
  no.~1, pp. 229--245, Feb. 2018.

\bibitem{xu2018energy}
L.~Xu and W.~Zhuang, ``Energy-efficient cross-layer resource allocation for
  heterogeneous wireless access,'' \emph{IEEE Trans on Wireless Commun.},
  vol.~17, no.~7, pp. 4819--4829, Jul. 2018.

\bibitem{Apollinaire2015A}
A.~Nadembega, A.~S. Hafid, and T.~Taleb, ``A destination and mobility path
  prediction scheme for mobile networks,'' \emph{IEEE Trans. Veh. Technol.},
  vol.~64, no.~6, pp. 2577--2590, Jun. 2015.

\bibitem{LSTM_highway}
F.~Altche and A.~de~La~Fortelle, ``An {LSTM} network for highway trajectory
  prediction,'' in \emph{Proc. IEEE ITSC Workshop}, 2017.

\bibitem{CGZ2018ICC}
C.-Y. Lin, K.-C. Chen, D.~Wickramasuriya, S.-Y. Lien, and R.~D. Gitlin,
  ``Anticipatory mobility management by big data analytics for ultra-low
  latency mobile networking,'' in \emph{Proc. IEEE ICC}, 2018.

\bibitem{bui2016anticipatory}
N.~Bui, M.~Cesana, S.~A. Hosseini \emph{et~al.}, ``A survey of anticipatory
  mobile networking: Context-based classification, prediction methodologies,
  and optimization techniques,'' \emph{IEEE Commun. Surveys Tuts.}, vol.~19,
  no.~3, pp. 1790--1821, Thirdquarter, 2017.

\bibitem{Hatem2014EE}
H.~Abou-zeid, H.~S. Hassanein, and S.~Valentin, ``Energy-efficient adaptive
  video transmission: Exploiting rate predictions in wireless networks,''
  \emph{IEEE Trans. Veh. Technol.}, vol.~63, no.~5, pp. 2013--2026, Jun. 2014.

\bibitem{Apollinaire2015Mobility}
A.~Nadembega, A.~Hafid, and T.~Taleb, ``Mobility-prediction-aware bandwidth
  reservation scheme for mobile networks,'' \emph{IEEE Trans. Veh. Technol.},
  vol.~64, no.~6, pp. 2561--2576, Jun. 2015.

\bibitem{DA2016ICC}
D.~Tsilimantos, A.~Nogales-G¡äomez, and S.~Valentin, ``Anticipatory radio
  resource management for mobile video streaming with linear programming,'' in
  \emph{Proc. IEEE ICC}, 2016.

\bibitem{Robert2014INFOCOM}
R.~Margolies, A.~Sridharan, V.~Aggarwal \emph{et~al.}, ``Exploiting mobility in
  proportional fair cellular scheduling: Measurements and algorithms,''
  \emph{IEEE/ACM Trans. Networking}, vol.~24, no.~1, pp. 355--367, Feb. 2016.

\bibitem{Yao2016Planning}
C.~Yao, C.~Yang, and Z.~Xiong, ``Energy-saving predictive resource planning and
  allocation,'' \emph{IEEE Trans. Commun.}, vol.~64, no.~12, pp. 5078--5095,
  Dec. 2016.

\bibitem{Ramy2017TWC}
R.~Atawia, H.~S. Hassanein, H.~Abou-zeid, and A.~Noureldin, ``Robust content
  delivery and uncertainty tracking in predictive wireless networks,''
  \emph{IEEE Trans. Wireless Commun.}, vol.~16, no.~4, pp. 2327--2339, Apr.
  2017.

\bibitem{Ramy2018GCT}
R.~Atawia, H.~S. Hassanein, N.~A. Ali, and A.~Noureldin, ``Utilization of
  stochastic modeling for green predictive video delivery under network
  uncertainties,'' \emph{IEEE Trans. Green Commun. and Netw.}, vol.~2, no.~2,
  pp. 556--569, June 2018.

\bibitem{guo2018icc}
C.~Yao, J.~Guo, and C.~Yang, ``Achieving high throughput with predictive
  resource allocation,'' in \emph{IEEE GlobalSIP}, 2016.

\bibitem{guo2018interference}
K.~Guo, T.~Liu, C.~Yang, and Z.~Xiong, ``Interference coordination and resource
  allocation planning with predicted average channel gains for {HetNets},''
  \emph{IEEE Access}, vol.~6, pp. 60\,137--60\,151, 2018.

\bibitem{Suhail2017ICC}
S.~Ahmad, R.~Reinhagen, L.~S. Muppirisetty, and H.~Wymeersch, ``Predictive
  resource allocation evaluation with real channel measurements,'' in
  \emph{Proc. IEEE ICC}, 2017.

\bibitem{ICSP18}
W.~Zhang, Y.~Liu, T.~Liu, and C.~Yang, ``Trajectory prediction with recurrent
  neural networks for predictive resource allocation,'' in \emph{Proc. IEEE
  ICSP}, 2018.

\bibitem{Radiomap}
J.~Chen, U.~Yatnalli, and D.~Gesbert, ``Learning radio maps for {UAV}-aided
  wireless networks: A segmented regression approach,'' in \emph{Proc. IEEE
  ICC}, 2017.

\bibitem{VTC18radiomap}
J.~Thrane, M.~Artuso, D.~Zibar, and H.~L. Christiansen, ``Drive test
  minimization using deep learning with bayesian approximation,'' in
  \emph{Proc. IEEE VTC Fall}, 2018.

\bibitem{A2014Scenarios}
{A. Osseiran, F. Boccardi and V. Braun, \emph{et al.}}, ``Scenarios for 5{G}
  mobile and wireless communications: The vision of the {METIS} project,''
  \emph{IEEE Commun. Mag}, vol.~52, no.~5, pp. 26--35, May. 2014.

\bibitem{3GPPQoS}
3GPP, \emph{Further Advancements for E-{UTRA} Physical Layer Aspects}.\hskip
  1em plus 0.5em minus 0.4em\relax TSG RAN TR 36.814 v9.0.0, Mar. 2010.

\bibitem{Juluri2015VoD}
P.~Juluri, V.~Tamarapalli, and D.~Medhi, ``Measurement of quality of experience
  of video-on-demand services: A survey,'' \emph{IEEE Commun. Surveys. Tuts.},
  vol.~18, no.~1, pp. 401--418, First quarter, 2016.

\bibitem{View20175GPPP}
{5GPPP Architecture Working Group}, ``View on 5{G} architecture,'' in \emph{5G
  Architecture White Paper}, Dec. 2017.

\bibitem{bera2017software}
S.~Bera, S.~Misra, and A.~V. Vasilakos, ``Software-defined networking for
  internet of things: A survey,'' \emph{IEEE Internet of Things J.}, vol.~4,
  no.~6, pp. 1994--2008, Dec. 2017.

\bibitem{ksentini2017toward}
A.~Ksentini and N.~Nikaein, ``Toward enforcing network slicing on {RAN}:
  Flexibility and resources abstraction,'' \emph{IEEE Commun. Mag.}, vol.~55,
  no.~6, pp. 102--108, Jun. 2017.

\bibitem{Seng2006}
S.~Wee-Seng and S.~K. Hyong, ``A predictive bandwidth reservation scheme using
  mobile positioning and road topology information,'' \emph{IEEE/ACM Trans.
  Netw.}, vol.~14, no.~5, pp. 1078--1091, Oct. 2006.

\bibitem{Hatem2014MSWiM}
H.~Abou-zeid, H.~S. Hassanein, and R.~Atawia, ``Towards mobility-aware
  predictive radio access: Modeling, simulation, and evaluation in {LTE}
  networks,'' in \emph{Proc. ACM MSWiM}, 2014.

\bibitem{VTC18CSIprediction}
W.~Jiang and H.~D. Schotten, ``Neural network-based channel prediction and its
  performance in multi-antenna systems,'' in \emph{Proc. IEEE VTC Fall}, 2018.

\bibitem{Apollinaire2016ICC}
A.~Nadembega, A.~S. Hafid, and R.~Brisebois, ``Mobility prediction model-based
  service migration procedure for follow me cloud to support {Q}o{S} and
  {Q}o{E},'' in \emph{IEEE ICC}, 2016.

\bibitem{Patrick2012Video}
P.~Seeling and M.~Reisslein, ``Video transport evaluation with {H}.264 video
  traces,'' \emph{IEEE Commun. Surveys Tuts.}, vol.~14, no.~4, pp. 1142--1165,
  2012.

\bibitem{Neely2015Adaptive}
D.~Bethanabhotla, G.~Caire, and M.~J. Neely, ``Adaptive video streaming for
  wireless networks with multiple users and helpers,'' \emph{IEEE Trans.
  Commun.}, vol.~63, no.~1, pp. 268--285, Jan. 2015.

\bibitem{Tang2007QoS}
J.~Tang and X.~Zhang, ``Quality-of-service driven power and rate adaptation for
  multichannel communications over wireless links,'' \emph{IEEE Trans. Wireless
  Commun.}, vol.~6, no.~12, pp. 4349--4360, Dec. 2007.

\bibitem{RALIU}
L.~Liu, P.~Parag, J.~Tang, W.-Y. Chen, and J.-F. Chamberland, ``Resource
  allocation and quality of service evaluation for wireless communication
  systems using fluid models,'' \emph{IEEE Trans. Inf. Theory}, vol.~53, no.~5,
  pp. 1767--1777, May 2007.

\bibitem{EB}
C.~Chang and J.~A. Thomas, ``Effective bandwidth in high-speed digital
  networks,'' \emph{IEEE J. Sel. Areas Commun.}, vol.~13, no.~6, pp.
  1091--1100, Aug. 1995.

\bibitem{EC}
D.~Wu and R.~Negi, ``Effective capacity: A wireless link model for support of
  quality of service,'' \emph{IEEE Trans. Wireless Commun.}, vol.~2, no.~4, pp.
  630--643, Jul. 2003.

\bibitem{Claude2012Flexible}
{C. Desset \emph{et al.}}, ``Flexible power modeling of {LTE} base stations,''
  in \emph{Proc. IEEE WCNC}, 2012.

\bibitem{earth2010}
{G. Auer, O. Blume, V. Giannini, I. G\'{o}dor, \emph{et al.}}, ``D 2.3: Energy
  efficiency analysis of the reference systems, areas of improvements and
  target breakdown,'' \emph{EARTH}, Jan. 2012.

\bibitem{she2016energy}
C.~She and C.~Yang, ``Energy efficiency and delay in wireless systems: Is their
  relation always a tradeoff?'' \emph{IEEE Trans. on Wireless Commun.},
  vol.~15, no.~11, pp. 7215--7228, Nov. 2016.

\bibitem{Riiser2012Video}
H.~Riiser, T.~Endestad, P.~Vigmostad, C.~Griwodz, and P.~Halvorsen, ``Video
  streaming using a location-based bandwidth-lookup service for bitrate
  planning,'' \emph{ACM Trans. Multimedia Comput. Commun. Appl.}, vol.~8,
  no.~3, pp. 24:1--24:19, Jul. 2012.

\bibitem{Xuan2015hotmobile}
X.~K. Zou, J.~Erman, V.~Gopalakrishnan, E.~Halepovic, R.~Jana, X.~Jin,
  J.~Rexford, and R.~K. Sinha, ``Can accurate predictions improve video
  streaming in cellular networks?'' in \emph{ACM HotMobile}, 2015.

\bibitem{Gregory2018Constrained}
J.~Gregory, \emph{Constrained Optimization In The Calculus Of Variations and
  Optimal Control Theory}.\hskip 1em plus 0.5em minus 0.4em\relax Chapman and
  Hall/CRC, 2018.

\bibitem{WirelessCom}
A.~Goldsmith, \emph{Wireless Communications}.\hskip 1em plus 0.5em minus
  0.4em\relax Cambridge University Press, 2005.

\bibitem{boyd}
S.~Boyd and L.~Vandanberghe, \emph{{C}onvex {O}ptimization}.\hskip 1em plus
  0.5em minus 0.4em\relax Cambridge Univ. Press, 2004.

\bibitem{VideoTrace}
\BIBentryALTinterwordspacing
{Video Trace Library}. [Online]. Available:
  \url{http://trace.eas.asu.edu/videotraces2/cgs/cif/Sony_G16B15_CIF_DQP6_5EL_48_42_36_30_24_18/}
\BIBentrySTDinterwordspacing

\bibitem{Claude2014Modeling}
C.~Desset, B.~Debaillie, and F.~Louagie, ``Modeling the hardware power
  consumption of large scale antenna systems,'' in \emph{Proc. IEEE GreenComm},
  2014.

\bibitem{TETC2013}
H.~A. Omar, W.~Zhang, A.~Abdrabou, and L.~Li, ``Performance evaluation of
  {V}e{MAC} supporting safety applications in vehicular networks,'' \emph{IEEE
  Trans. Emerging Topics in Computing}, vol.~1, no.~1, pp. 69--83, Jun. 2013.

\bibitem{Aharon2009Robust}
A.~Ben-Tal, L.~El~Ghaoui, and A.~Nemirovski, \emph{Robust Optimization}.\hskip
  1em plus 0.5em minus 0.4em\relax Princeton University Press, 2009.

\end{thebibliography}

\end{document}